\documentclass[a4paper,11pt]{article}
\usepackage{jinstpub} % for details on the use of the package, please see the JINST-author-manual
\usepackage{lineno}
\usepackage{subcaption} 
\usepackage{caption} 
\usepackage{multirow}
\usepackage{mathtools}     % needed for "dcases*"

\usepackage{array}
\title{\boldmath Numerical Modelling of Active Target Time Projection Chamber for Low Energy Nuclear Physics}
\author[a]{Pralay Kumar Das,}
\affiliation[a]{Saha Institute of Nuclear Physics,  a CI  of Homi Bhabha National
Institute, AF Block, Sector I, Bidhannagar, Kolkata - 700064, India}
\emailAdd{pralay.das@saha.ac.in}
\author[b]{Jaydeep Datta,}
\affiliation[b]{Center for Frontiers in Nuclear Science, Department of Physics and Astronomy, Stony
Brook University, 100 Nicolls Road, Stony Brook, New York-11794, USA}
\author[a]{Nayana Majumdar}
\author[a,*]{and Supratik Mukhopadhyay}

\abstract{A numerical model based on hydrodynamic approach has been developed to emulate the device dynamics of active target Time Projection Chamber which is utilized for studying nuclear reaction through three dimensional tracking of concerned low energy particles. The proposed model has been used to investigate the performance of a prototype active target Time Projection Chamber, namely SAT-TPC, to be fabricated at Saha Institute of Nuclear Physics, for its application in nuclear physics experiments. A case study of non-relativistic elastic scattering $^4He+^{12}C$ with beam energy $25~MeV$ and current $2.3~pA$ has been opted for this purpose. The effect of beam induced space charge on the tracking performance the SAT-TPC prototype has been studied to optimize the beam current and scheme of the anode readout segmentation. The model has been validated by comparing its results to that of a particle model used to explain observed distortion in scattered particle tracks in a low energy nuclear physics experiment.}


\begin{document}
\maketitle
\flushbottom
% \begin{frontmatter}

\section{Introduction}
\label{introduction}
The Time Projection Chamber (TPC) \cite{Nyg1974,Mar1978} serves as a single compact device which can offer a novel scheme for reconstructing three dimensional particle trajectories owing to its configuration and working principle. The device consists of a large volume of gas, maintained under a constant electric field. It allows primary electrons, produced in the gaseous medium by the passage of ionizing charged particle, to drift over a large distance following the electric field lines. The electrons are subsequently led to a position sensitive multiplier stage placed at the end of the drift volume, equipped with suitably high electric field configuration and segmented anode plane. At the multiplier stage, the primary electrons undergo further ionization processes under the action of the local electric field and amplify in number, giving rise to an avalanche. Eventually, the two dimensional (2D) position information of the primary event is obtained from the signal induced by the electronic avalanche on the readout segments of the anode. The time of arrival of the electrons is utilized to determine their drifting distance from their drift velocity in the given gas medium which finally provides the third position information of the event. 

In addition to its  three dimensional (3D) tracking ability, the use of magnetic field, set parallel to the electric field of the TPC to minimize the transverse diffusion of the electrons that degrades the position resolution of the device, has substantiated the application of the device in collider based high energy physics experiments. The combination of handling high particle flux and large track densities due to low occupancy has paved its way as preferred particle tracker over a fairly long period since its introduction \cite{Att2009,Hil2010,Die2018}. In due course of time, the TPC has undergone many improvisations and followed implementation eventually in several other areas of research in fundamental physics, such as rare event search and low energy nuclear physics, where three dimensional tracking of various product particles and their identification play important roles \cite{Die2018}. 

A unique implementation of the TPC in low energy nuclear physics experiments has emerged with the advent of radioactive beams by using its active gaseous medium as the target of a nuclear reaction. The device, coined as Active Target Time Projection Chamber (AT-TPC), may offer several notable advantages for studying low energy nuclear reactions in comparison to traditional experimental setups, involving solid targets and detector arrays. While performing some nuclear reactions with radioactive beams using inverse kinematics, the energies of the recoil particles can vary widely. The gaseous target eliminates the loss of energy within the target material which becomes critical in studying low energy products. In addition, the low intensities of these beams substantially limit the luminosity of the experiments which employ solid targets. A difficult compromise between luminosity and resolution is called for to mitigate the limitation by increasing the target thickness. AT-TPC can directly address this issue by using the gaseous target as the detector medium simultaneously. Therefore, the increase in target thickness as well as luminosity can be implemented without impacting the energy and angular resolutions. Furthermore, the configuration of the device offers $4\pi$ coverage in acceptance and three dimensional tracking of projectile and reaction products including those emitted with very low kinetic energy, and thus facilitate recording of all information, such as vertex, energies etc.  Due to a generally large gas thickness crossed by the ionising particle, the device offers a fair capability of particle identification as well through energy deposition. The basic implementation of  AT-TPC is thus conceptually similar to that usually done with collider based TPC, however at much smaller scale of physical dimension. Likewise many other applications, the AT-TPC configuration too has undergone many adjustments in order to address different problems of low energy nuclear physics \cite{Ayy2018,Baz2020}. 

There are several issues as well in context to the application of the AT-TPC for studying nuclear reactions. The performance of the device is governed by a few factors in which gaseous medium and electric field distribution are of crucial importance. The choice of active gas and its pressure in AT-TPC needs special attention in order to ensure its suitability for adequate interaction probability and stopping power of concerned nuclear reaction products.  Consequently, it calls for optimization of the position sensitive multiplier stage for its performance in the active gas and pressure. A compromise between purity of the active gas and the electronic gain in the position sensitive multiplier stage is unavoidable in order to achieve satisfactory position resolution for precise tracking of the particles. Simultaneously, it gives rise to additional unwanted reaction channels which requires selective trigger, suitably designed either using external ancillary detectors or internally with special electronics. Another issue concerns the space charge produced by the low energy particles in the active gas medium through ionization. The space charge density may be substantial depending upon the beam rate and luminosity to cause local distortion of the electric field distribution which should be otherwise homogeneous to achieve precise timing information useful for extracting the third dimensional position information of the event. The effect of beam induced space charge on scattered particle and beam tracks has been observed experimentally \cite{Ran2019}. 
There exists another possibility of perturbation of the electric field near the multiplier stage due to a process, termed as Ion Back Flow (IBF) where the ions produced in the multiplier stage enter the drift volume to infiltrate the uniform field \cite{deb2017,Purba2017}. Although the issue of IBF is likely to be substantial in high flux collider based experiments, its effect in AT-TPC too needs attention as this physical process may influence the tracking performance of the device. However, detailed experimental or numerical study on this aspect is still scarce.  

In the present work, a numerical model based on hydrodynamic approach has been reported which introduces a scheme of emulating the device dynamics of the AT-TPC. The same model has been implemented earlier to study a few important aspects related to device dynamics of Resistive Plate Chamber (RPC) \cite{Dat2020,Dat2021} and Gaseous Electron Multiplier (GEM) \cite{Rou2021,Rou2021disprob}. The efficacy of the proposed hydrodynamic model for the present application has been investigated by comparing its results to that produced by the particle model used in \cite{Ran2019} where the transport of individual electrons has been considered to provide a qualitative explanation of the observed effects in their experiment. It may be worth mentioning in this context that being an Eulerian approach, the hydrodynamic model employs less computational resource in comparison to the electron transport model which is basically a Lagrangian method. The present model has been further used to conduct a numerical test of an AT-TPC, namely Saha-AT-TPC (SAT-TPC), planned to be fabricated at Saha Institute of Nuclear Physics, Kolkata, for realizing its operational and tracking performance limits. The test has been conducted to study the performance of the SAT-TPC in tracking the reaction/scattered products of a specific case of non-relativistic elastic scattering $^4He$ + $^{12}C$ and the effect of beam induced space charge on it. The optimization of the anode readout segmentation for tracking the reaction products with the given space charge effect has been performed using the model.

The outline of the present article is the following. In section \ref{NumMod}, a brief description of the numerical model, based on hydrodynamics, has been provided with all of its modules/stages discussed individually. It is followed by the section \ref{ModVal} where the validation of the numerical model has been discussed with relevant results. The next section \ref{SATMod} has presented the numerical modelling of the SAT-TPC for studying its performance for a specific case of the non-relativistic elastic scattering.  The beam induced space charge and its effect on tracking have been reported in section \ref{BeaSpa}, followed by relevant optimization studies on beam current and anode readout segmentation. Finally, the article has ended with section \ref{SumCon}, presenting the summary and conclusion of the present work. 

\section{Numerical Model}
\label{NumMod}
\begin{figure}[h!]
 \centering
     \includegraphics[scale=0.25]{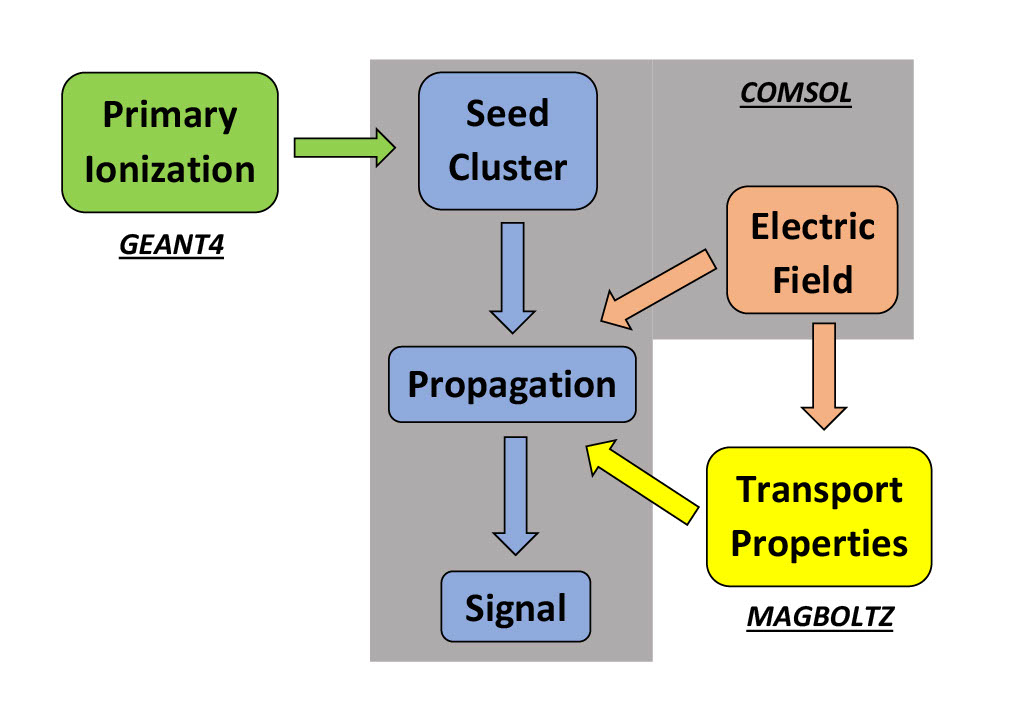}
 \caption{Flowchart of the numerical model}
     \label{flowchart}
   \end{figure}

The proposed numerical model utilizes the hydrodynamic principle where the movement of the negative and positively charged particles in the gas volume has been considered to be analogous to the mass transport of dissolved solute species in a solvent. The numerical model has been built on COMSOL Multiphysics \cite{Com} to simulate the propagation of the charged species (ions, electrons) in the drift volume of the AT-TPC in presence of the applied electric field. The packages based on the particle model, like GEANT4 \cite{Gea2003} and MAGBOLTZ \cite{Mag}, have been used as supporting toolkits for producing necessary data on primary ionization and transport properties of the charged species in the given gas medium respectively, while the electric field has been computed on the COMSOL platform itself. For the present work, no multiplier stage has been considered in the AT-TPC geometry and hence no amplification of the primary electrons. A schematic flowchart of the entire numerical model has been illustrated in figure \ref{flowchart}. All the modules/stages of the numerical model have been briefly discussed below.

\subsection{Primary Ionization}
The primary ionization of the gas molecules, caused by a specific low energy charged particle along its trajectory within the drift volume of the AT-TPC, has been simulated using physics lists of GEANT4, including Penelope, Livermore, and Photo Absorption and Ionization (PAI) that incorporates low energy electromagnetic processes. 

\subsection{Seed Cluster}
The spatial distribution of the primary charges have been used to produce a parametric definition of the distribution. For this purpose, the spatial distributions along $X, Y, Z$- directions have been fitted with suitable mathematical functions. The fitting parameters have been used to configure a seed cluster of the relevant charged species. 

\subsection{Electric Field}
The simulation of electric field configuration of the device has been done by building the AT-TPC geometry in COMSOL which has been suitably meshed to achieve results at reasonable computational expense.  The electric field has been computed with the {\it{Electrostatic}} module which solves Poisson's equation using Finite Element Method (FEM) and the electric field, $\vec{E}$ using the scalar electric potential, $V$, as the dependent variable (vide equations \ref{eq1} - \ref{eq2}). 
\begin{equation}
	{\nabla}^2{V}=-\rho/\epsilon_0
	\label{eq1}
\end{equation} 
\begin{equation}
 	\vec{E}=-\vec{\nabla}V
	\label{eq2}
\end{equation}
Here, $\rho$ and $\epsilon_0$ denote the charge density and permittivity of space, respectively.

\subsection{Transport Properties}
The transport properties of the primary electrons in the given gaseous medium of specific pressure and temperature and for the applied electric field have been computed using MAGBOLTZ. The drift velocity and the diffusion coefficients (longitudinal and transverse) are the properties of relevance that have been useful for emulating the propagation of the seed cluster. However, the MAGBOLTZ can not deliver the transport properties of the ions due to non-availability of adequate experimental data and theoretical substantiation.

\subsection{Propagation}
In the hydrodynamic model, the active gas medium has been considered as a charged solution with dilute concentration. The neutral gas molecules have been considered as the solvent while the charged solutes are electrons and ions, produced in the medium by the processes of primary ionization. The solution has been considered dilute as the concentration of the charged species is small in comparison to that of the neutral molecules. The {\it{Transport of Dilute Species (TDS)}} module of the COMSOL package has been used to emulate the field dependent propagation of the seed cluster by solving the following drift diffusion reaction equation \ref{eq3}. 
\begin{equation}
\frac{\partial c_{j}}{\partial t} + \vec{\nabla}(-D_j\vec{\nabla} c_j+\vec{u_j} c_j)=S_j
\label{eq3}
\end{equation}
where $c_{j}$,$D_{j}$, $\vec{u_j}$ and $S_{j}$ in equation \ref{eq3} denote the concentration, rate of production, drift velocity, and diffusion coefficients for the charged species ($j$ = ion, electron), present in the seed cluster. 

\subsection{Signal}
The electronic signal, produced on the segmented anode, placed at the end of the drift volume, has been calculated from the amount of electronic charge flowing through a given area following the expression given below.
\begin{equation}
	i(t) = q_e.n_e(t).\vec{u_e}(t).A
	\label{eq4}
\end{equation}  
where $i(t)$, $q_e$, $n_e(t)$, $\vec{u_e}(t)$ are the electronic current, charge, number density and drift velocity while flowing through an area $A$ on the anode at an instant $t$.

\section{Model Validation}
\label{ModVal}
For validation of the proposed numerical model, it has been implemented to simulate the beam induced space charge effect on scattered particle track, as observed in a low energy nuclear physics experiment using AT-TPC \cite{Ran2019}. There, a detailed particle model with electron transport simulation has been performed using Garfield \cite{veenhof1998} to provide qualitative explanation of the observed distortions in the scattered particle tracks. 

\begin{figure}[h!]
     \centering
     \begin{subfigure}[b]{0.45\linewidth}
         \centering
         \includegraphics[width=1.1\textwidth]{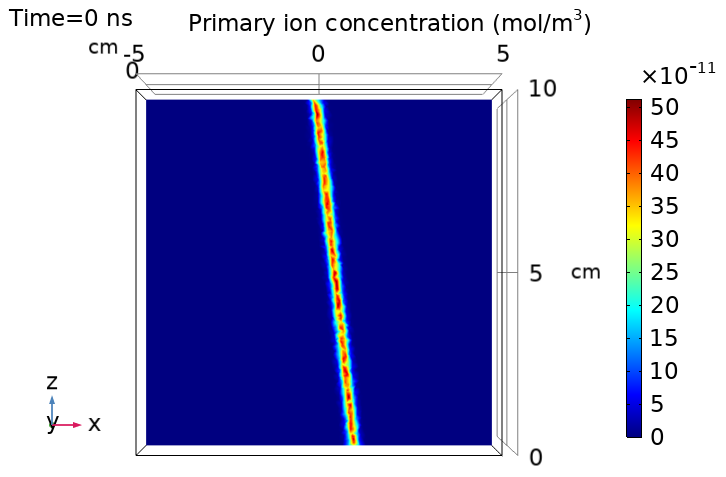}
         \caption{}
         \label{sp_charge}
     \end{subfigure}
     \hspace{10mm}
     \begin{subfigure}[b]{0.45\linewidth}
         \centering
         \includegraphics[width=0.9\textwidth]{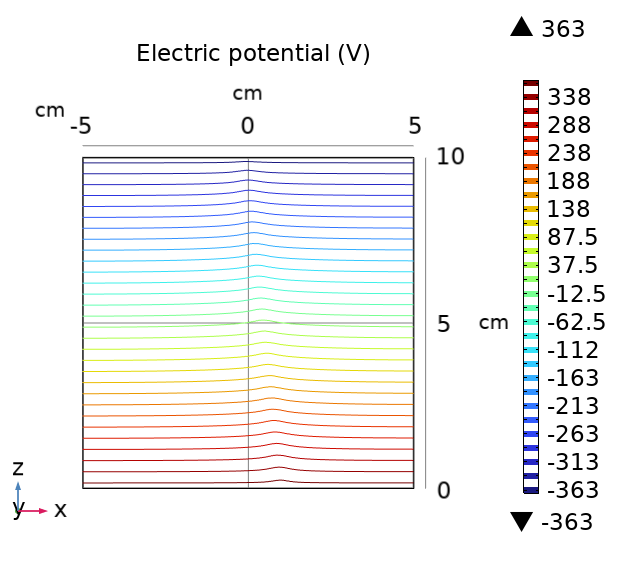}
         \caption{}
         \label{pot_dis}
     \end{subfigure}
        \caption{(a) Space charge density along the beam direction, and (b) potential distribution in presence of space charge}
        \label{fig:model}
\end{figure}

For reduced computation, a scaled down geometry of the AT-TPC as mentioned in \cite{Ran2019}, with height $10~cm$ (along $Z$-direction) and cross section $5~cm \times 5~cm$ (in $XY$-plane), has been modeled in COMSOL. The same uniform electric field of magnitude $75~V/cm$, as mentioned in \cite{Ran2019}, has been produced along the central $Z$-axis. Following the experimental scenario, a tilted configuration of the device ($6.2^\circ$ with respect to $Z$-axis) has been considered. As per the referred work, a space charge density $10^{-4}~C/m^3$ has been distributed in a Gaussian profile with a standard deviation of $1~mm$ to represent the ionization caused by $^{46}K$ beam of energy $4.6~MeV/u$ and instantaneous rate $10^4~pps$ (particle per second) in P10 gas, kept at $100~Torr$ to fill the AT-TPC. In figure \ref{sp_charge}, the space charge distribution along the beam direction in the active volume of the AT-TPC has been illustrated. The potential distribution, as computed by the COMSOL with the given space charge density, has been depicted in figure \ref{pot_dis} where distortion around and along the beam direction can be seen which is similar to the plot presented in the referred work.  

\begin{figure}[htbp]
\centering
\includegraphics[width=.75\textwidth]{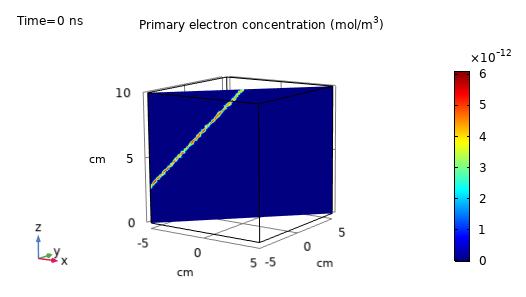}
\caption{ Scattered particle track in the present model} 
\label{fig:scatt_track}
\end{figure}

Similar to a scattered particle track produced from the interaction of the $^{46}K$ beam with the P10 gas, as considered in the referred work, a distribution of primary electrons in a Gaussian profile with standard deviation $2~mm$ has been used in the present work. The same has been shown in figure \ref{fig:scatt_track}. This has been considered as the seed cluster of electrons to be propagated through the AT-TPC drift volume having its potential distribution as shown in figure \ref{pot_dis}, influenced by the $^{46}K$ beam induced space charge. 

\begin{figure}
     \centering
     \begin{subfigure}[b]{0.45\linewidth}
         \centering
         \includegraphics[width=\textwidth]{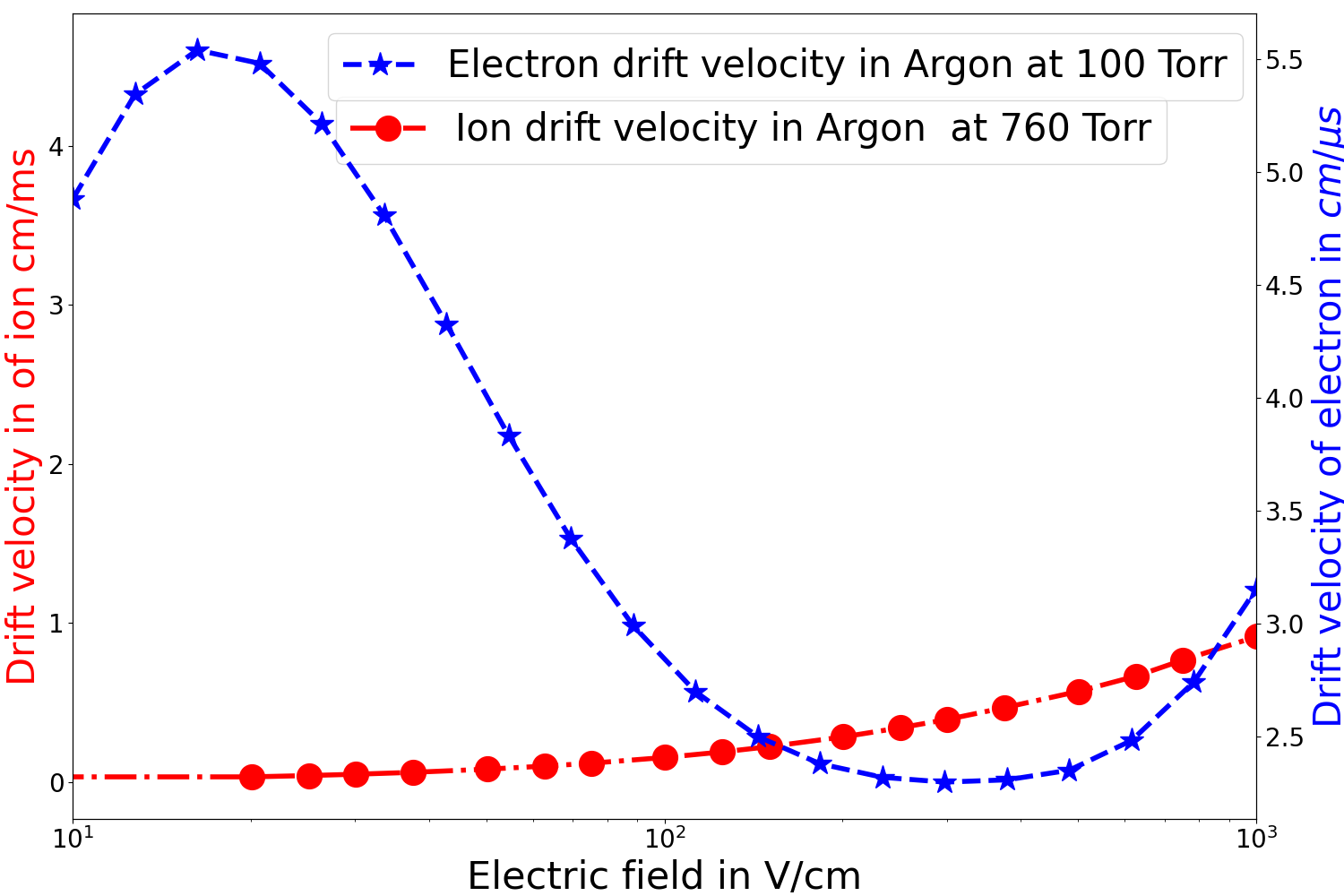}
         \caption{}
         \label{fig:drift}
     \end{subfigure}
     % \hspace{30 mm}
     \begin{subfigure}[b]{0.45\linewidth}
         \centering
         \includegraphics[width=\textwidth]{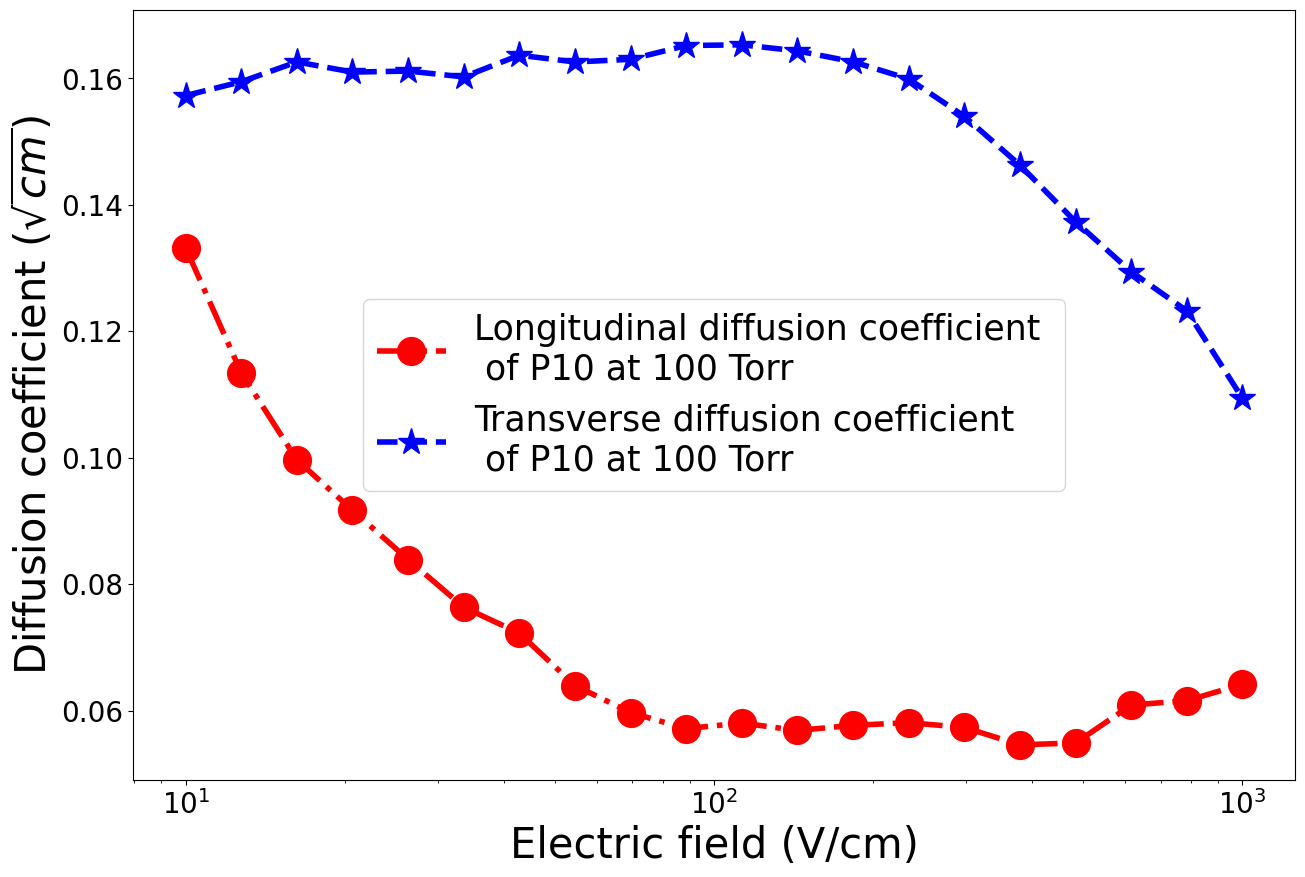}
         \caption{}
         \label{fig:dff}
     \end{subfigure}
   \caption{(a) Drift velocity of electrons (from MAGBOLTZ) and ions following \cite{Bas2000}, and (b) diffusion coefficients of electrons in P10 gas at 100 Torr (from MAGBOLTZ)}
        \label{fig:P10}
\end{figure}

The transport properties of the electrons in P10 gas at pressure $100~Torr$ have been determined using the MAGBOLTZ. The plots of the drift velocity and the diffusion coefficients as a function of electric field have been shown in figure \ref{fig:P10}. On the other hand, the drift velocity of the Argon ion in pressure $760~Torr$, as available from \cite{Bas2000} and plotted in figure \ref{fig:drift}, has been used for computing the propagation of ions present along the beam track since the data for $100~Torr$ is not available.

\begin{figure}[h!]
        \centering
        \begin{subfigure}[b]{0.45\linewidth}
            \centering
            \includegraphics[scale=0.3]{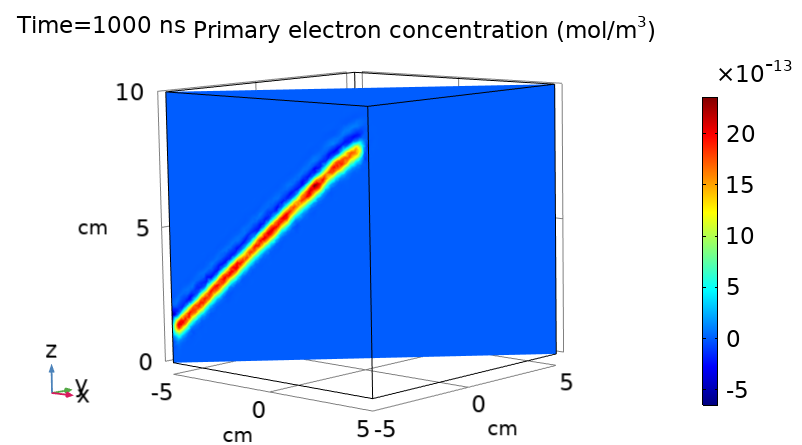}
            \caption[Network2]%
            {{}}    
            \label{fig:mean and std of net14}
        \end{subfigure}
        % \hspace{40 mm}
        \begin{subfigure}[b]{0.45\linewidth}  
            \centering 
            \includegraphics[scale=0.3]{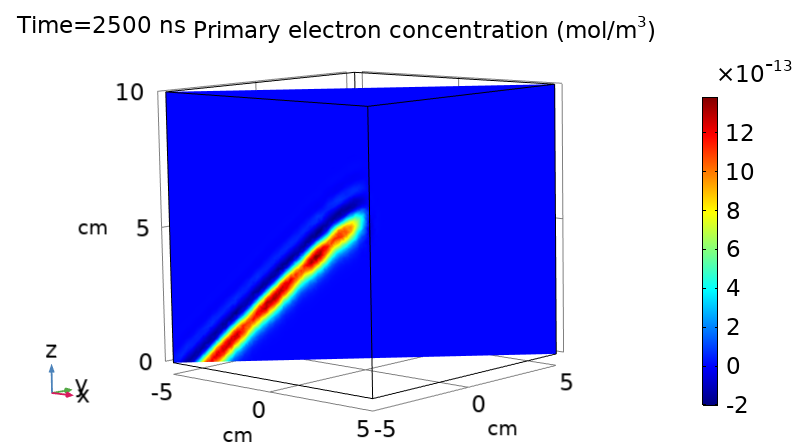}
            \caption[]%
            {{}}    
            \label{fig:mean and std of net24}
        \end{subfigure}
        \vskip\baselineskip
        \begin{subfigure}[b]{0.45\linewidth}   
            \centering 
            \includegraphics[scale=0.3]{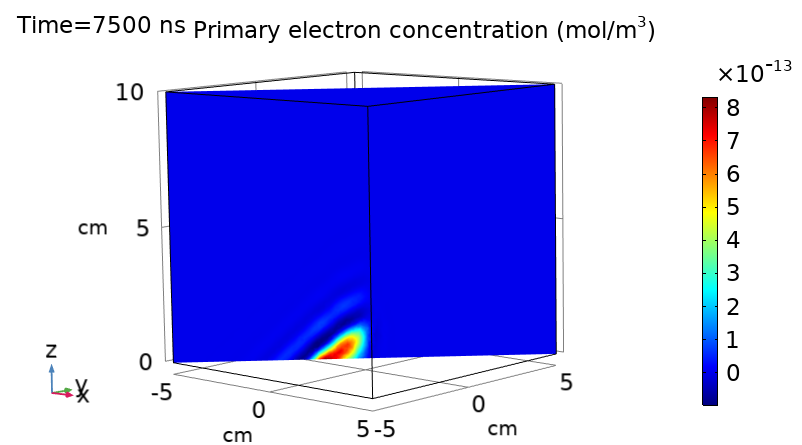}
            \caption[]%
            {{}}    
            \label{fig:mean and std of net34}
        \end{subfigure}
       % \hspace{40 mm}
        \begin{subfigure}[b]{0.45\linewidth}   
            \centering 
            \includegraphics[scale=0.3]{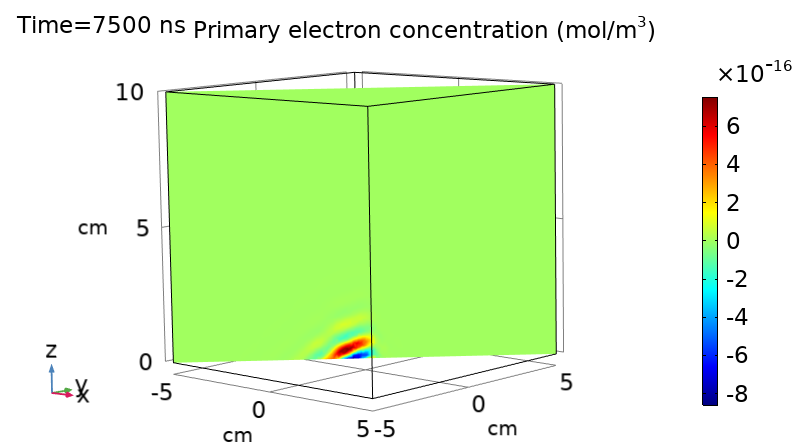}
            \caption[]%
            {{}}    
            \label{fig:mean and std of net44}
        \end{subfigure}
        \caption
        {Temporal evolution of primary electron concentration in $XZ$-plane at (a) $Time =1000~ns$, (b) $Time = 2500~ns$, (c) $ Time = 5000~ns$, and (d) $Time =7500~ns$} 
        \label{fig:mean and std of nets}
    \end{figure}

The propagation of the seed cluster of the electrons, distributed along the scattered track, through the space charge affected field has been simulated with the hydrodynamic model. The primary electron concentration has been recorded at a time step of $25~ns$ using the updated electric field and the corresponding transport parameters. The field at each time step has been calculated considering the temporal evolution of the ion concentration using their drift velocity obtained from \cite{Bas2000}. However, the mobility of the ionic space charge is quite low to produce a substantially dynamic electric field and hence the change in the original space charge affected electric field is nominal. A few cases of the temporal evolution of primary electron concentrations at several instants ($Time = 1000, 2500, 5000, 7500~ns$) in $XZ$-plane have been illustrated in figure \ref{fig:mean and std of nets}. It can be seen from the plots that part of the scattered track near the beam has undergone a bend in shape as well as reduction in concentration due to the distortion in the electric field caused by the space charge. The result has shown an agreement to the characteristic {\it{"knee"}} at the beginning of the tracks of the scattered particles / reaction products as was found experimentally and later reproduced with the particle model in \cite{Ran2019}.

\begin{figure}[h!]
     \centering
         \includegraphics[scale=0.35]{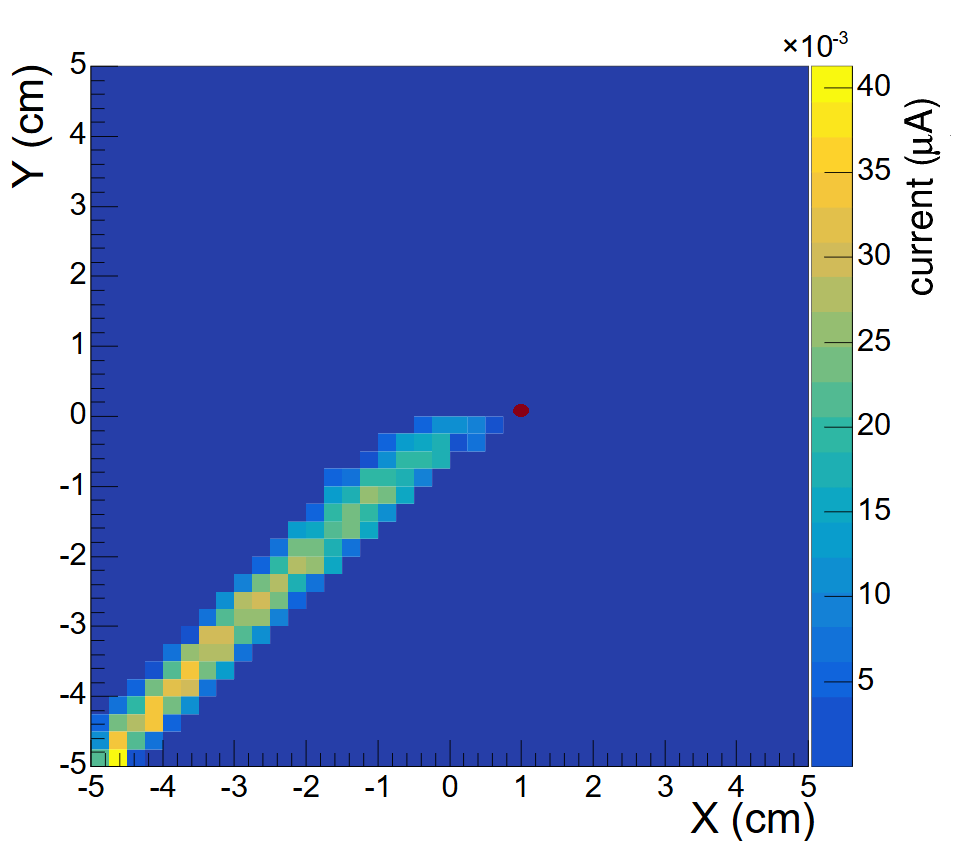}
        \caption{Current signal, produced by the scattered particle track cluster on the anode plane, segmented in $2~mm\times 2~mm$ pads, and the beam spot, marked by red circle}
        \label{fig:graph}
\end{figure}

The calculation has been furthered to observe the projected scattered particle track in terms of the current produced on the anode plane. For this purpose, the anode plane has been segmented in $2~mm\times 2~mm$ pads. In figure \ref{fig:graph}, the 3D histogram plot of the current signal produced on the anode plane has been illustrated where the distortion of the scattered particle track caused by the beam induced space charge is clearly visible near the beam spot, marked by a red circle.

\section{SAT-TPC Modelling}
\label{SATMod}
\begin{figure}[h!]
 \centering
     \includegraphics[scale=0.6]{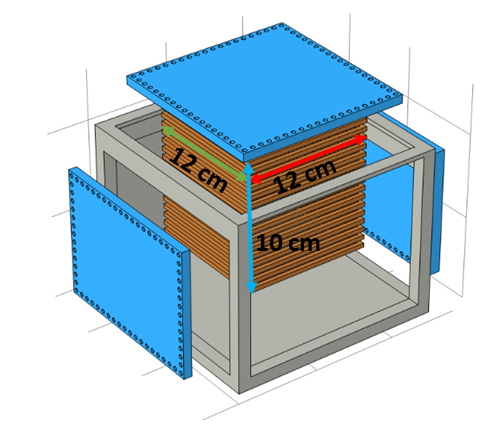}
 \caption{Schematic design of the prototype SAT-TPC}
     \label{ED}
   \end{figure}

\begin{figure}[h!]
     \centering
          \begin{subfigure}[b]{0.45\linewidth}
         \centering
         \includegraphics[scale=0.4]{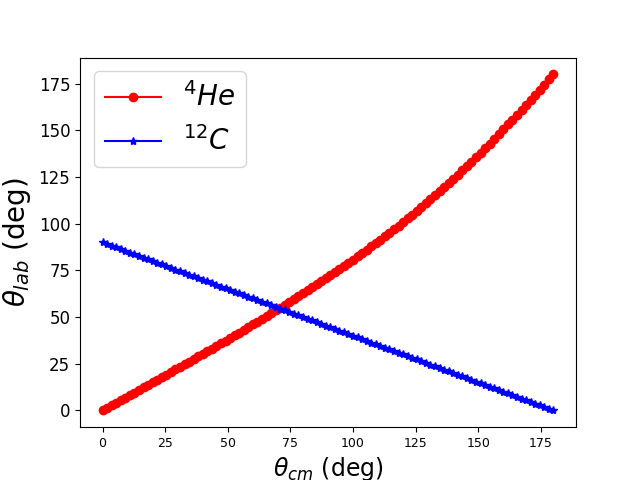}
         \caption{}
         \label{angle}
     \end{subfigure}
     % \hspace{40 mm}
     \begin{subfigure}[b]{0.45\linewidth}
         \centering
         \includegraphics[scale=0.4]{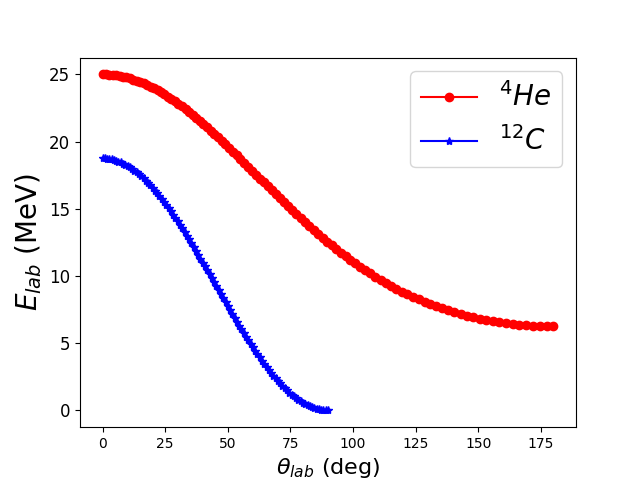}
         \caption{}
         \label{labEA}
     \end{subfigure}
        \caption{(a) Correlation of scattering angles in CM and lab frames, and (b) energy as a function of angle in lab frame of the scattered products $^4He$ and $^{12}C$}
\end{figure}

The performance of the proposed prototype SAT-TPC with the given design parameters, as shown in figure \ref{ED}, has been studied with the numerical hydrodynamic model to study its tracking capabilities. The evaluation of the effect of space charge, generated by the projectile beam on the scattered product tracks, has been the main objective of the current investigation. The optimization of the projectile beam rate  / current along with the anode segmentation has been exercised in this context on the basis of the tracking performance of the SAT-TPC.
The elastic scattering reaction $^{4}He$+$^{12}C$ has been used as a case study where the projectile $^4He$ interacts with the $^{12}C$ nuclei of the active gas mixture $Ar+CO_2$ of volumetric ratio 90:10. The two body kinematics of the non-relativistic elastic scattering $^4He+^{12}C$ has been solved for $^{4}$He projectile of energy $25~MeV$, bombarding on stationary $^{12}C$ target, present in the active gas mixture, for all possible angles in the centre of mass (CM) frame. The scattering angle and the corresponding energy of the reaction products $^{4}He$ and $^{12}C$ have been obtained and transformed to laboratory (lab) frame. The correlation of the CM and lab angles of the scattered products $^{4}He$ and $^{12}C$ has been plotted in figure \ref{angle} while the energy of the products as a function of scattering angle in the lab frame displayed in figure \ref{labEA}. For the limited scope of presentation, a specific case of the elastic scattering has been opted in the current study where the range of the scattered products is contained within the active volume. The angle and energy of the products in the lab frame of the case study have been quoted in table \ref{tab:my_label}.

 \begin{table}[h!]
    \centering
        \caption{The lab angle and energy of the scattered products $^4He$ and $^{12}C$ in the case study of elastic scattering $^4He + ^{12}C$}
\begin{tabular}{ | m{1.4cm} | m{2cm} | m{1.4cm} | m{2cm} |} 
  \hline
  $\theta^{lab}_{^{4}He} (deg)$  & $E^{lab}_{^4He} (MeV)$ & $\theta^{lab}_{^{12}C} (deg)$ & $E^{lab}_{^{12}C} (MeV)$\\ 
  \hline
 135 & 7.6 & 15.68 & 17.4 \\ 
  \hline
  \end{tabular}
    \label{tab:my_label}
\end{table}

\subsection{SAT-TPC Geometry}
A cuboid geometry of $1440~cm^3$ ($10~cm\times12~cm\times12~cm$) has been modeled in COMSOL to represent the drift volume of the SAT-TPC. The choice of dimension of the drift volume has been mainly governed by the available computational resource combined with low cost production of the prototype device. A comparison of the computation time of transport of electron species in the given $500~V/cm$ electric field for drift volumes obtained with $1.2~Teraflops$ computation power has been shown in table \ref{cp} to present the involved computational expense. A potential difference of $5000~V$ has been generated along the $Z$-axis of the drift volume by applying a voltage of $-5000~V$ to the cathode (at $Z = 10~cm$) while keeping the anode at ground (at $Z = 0~cm$). The anode plane ($12~cm\times12~cm$) has been segmented into square elements for reading out the collected signal. 

\begin{table}[h!]
    \caption{Computation time for different drift volume dimensions}
\begin{center}
\begin{tabular}{ | m{4cm} | m{4cm}| } 
  \hline
  Drift volume  & Computation time \\ 
  \hline
  $10~cm\times12~cm\times 12~cm $& 2 hrs 31 mins \\ 
  \hline
  $20~cm\times20~cm\times 20~cm$ &  4 hrs 39 mins \\ 
  \hline
\end{tabular}
\end{center}
    \label{cp}
\end{table}

\subsection{Primary Ionization}
\label{PriIon}
\begin{figure}[h!]
 \centering
     \includegraphics[scale=0.35]{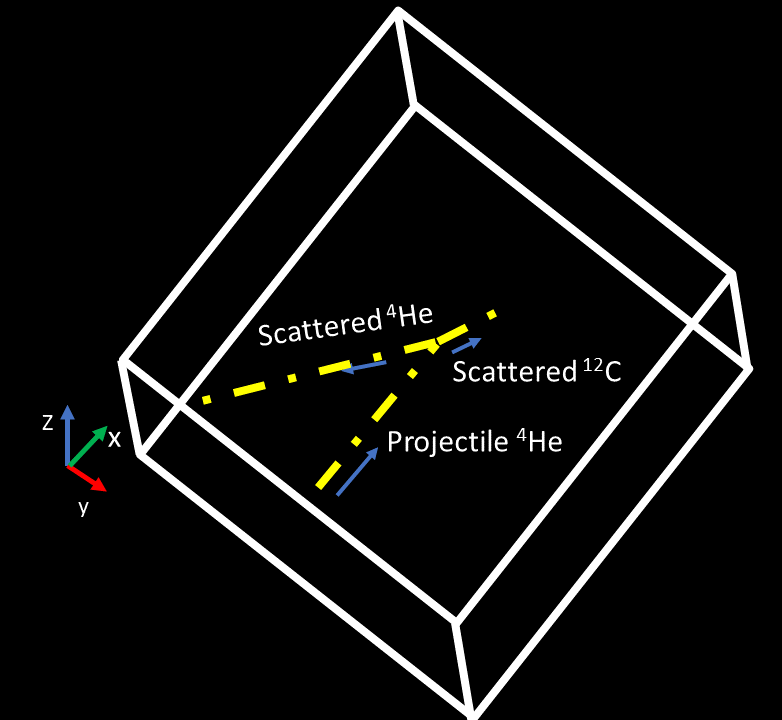}
 \caption{Schematic diagram of the specific elastic scattering case}
     \label{schem}
   \end{figure}

A schematic diagram has been shown in figure \ref{schem} to facilitate the visualization of the specific elastic scattering event to be studied by the prototype SAT-TPC as a numerical test case in the present work. It can be noted  that the beam of  $^4He$ of energy $25~MeV$ has been projected along the central $X$-axis. The elastic scattering event, as mentioned in table \ref{tab:my_label}, has been considered to occur at the centre of the device. Accordingly, the trajectories of the scattered products, $^4He$ and $^{12}C$, emitted with energies $7.6~MeV$, $17.4~MeV$, at angles $135^\circ$ and $15.68^\circ$ respectively with respect to the beam axis, have been depicted.
The calculation of primary ionization caused by the projectile $^4He$ and the scattered products $^4He$ and $^{12}C$ have been discussed in the following sub-sections.

\subsubsection{Projectile $^4He$}
\begin{figure}[htb]
     \centering
     \begin{subfigure}[b]{0.45\linewidth}
         \centering
         \includegraphics[width=\textwidth]{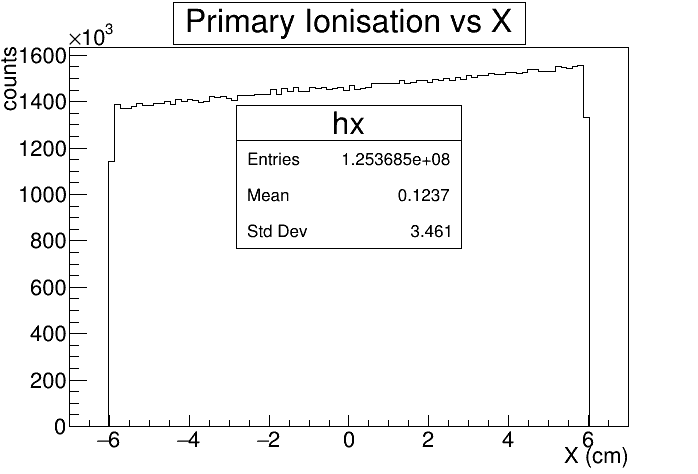}
         \caption{}
         \label{100_x}
     \end{subfigure}
     %\hfill
          % \hspace{10 mm}
     \begin{subfigure}[b]{0.45\linewidth}
         \centering
	\includegraphics[width=\textwidth]{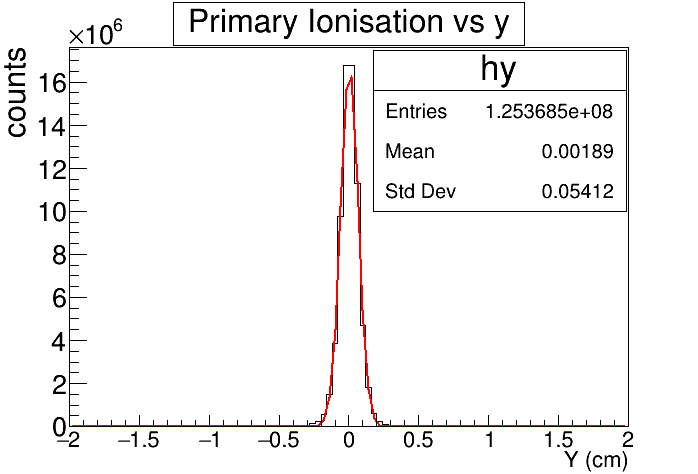}
         \caption{}
         \label{100_y}
     \end{subfigure}
     % \hspace{20 mm}
     \begin{subfigure}[b]{0.45\linewidth}
        \centering
         \includegraphics[width=\textwidth]{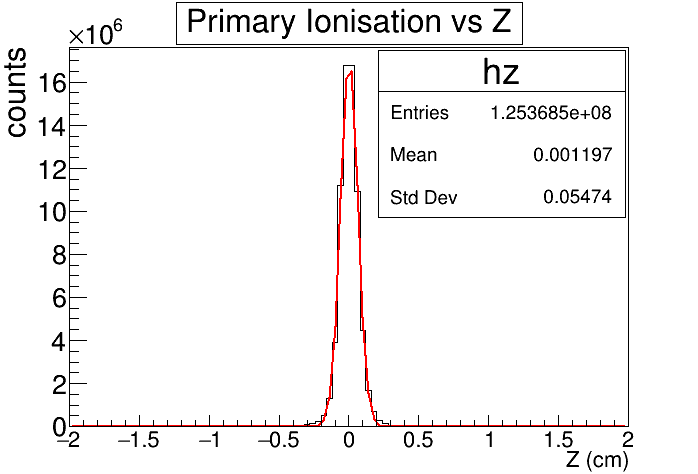}
         \caption{}
         \label{100_z}
     \end{subfigure}
        \caption{Primary ionization by $1000$ numbers of $^4He$ projectiles of energy $25~MeV$ along (a) $X$, (b) $Y$, and (c) $Z$-directions in the drift volume of SAT-TPC}
        \label{p_z}
\end{figure}

The number of primary charges, produced by $^4He$ projectile particle of energy $25~MeV$, has been estimated from the statistical average of $1000$ events obtained by using GEANT4. It has partially deposited its energy  while passing through the active volume of the TPC. The number of primary charges produced along the $X$-direction has been shown in figure \ref{100_x}. It has been approximated as a uniform distribution while the $Y$ and $Z$-distributions have been fitted with the Gaussian function, as shown in figures \ref{100_y} and \ref{100_z}.

The fitting parameters have been eventually utilized to form a positive ion distribution along the beam direction for computing the effect of space charge on the the electric field distribution in the SAT-TPC. The average number of primaries ionized by each beam particle has been found to be $\sim 1.25\times10^5$.

\subsubsection{Scattered Products $^4He$ and $^{12}C$}
\begin{figure}[h!]
     \centering
     \begin{subfigure}[b]{0.45\linewidth}
         \centering
         \includegraphics[scale=0.4]{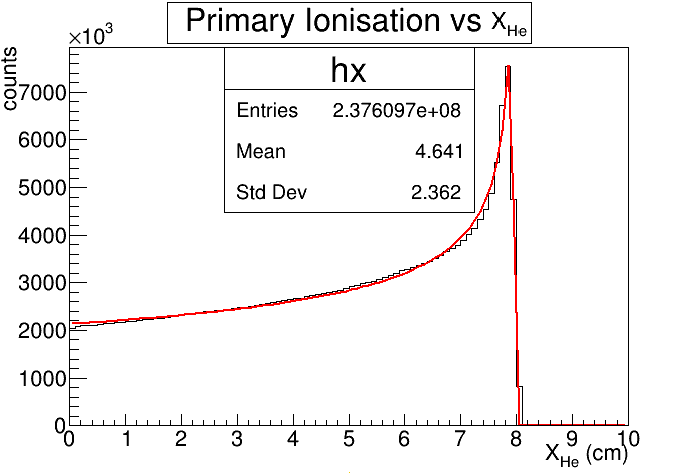}
         \caption{}
         \label{fig1:}
     \end{subfigure}
     % \hspace{25mm}
\begin{subfigure}[b]{0.45\linewidth}
         \centering
         \includegraphics[scale=0.4]{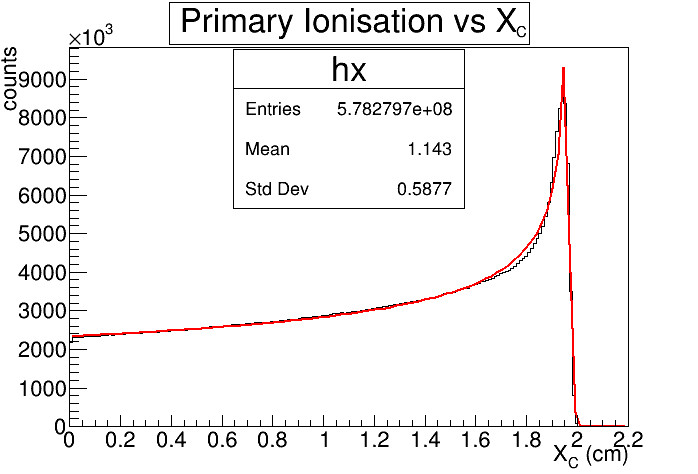}
         \caption{}
         \label{fig1:}
     \end{subfigure}
     \hfill
     % \hspace{20mm}
     \begin{subfigure}[b]{0.45\linewidth}
         \centering
         \includegraphics[scale=0.4]{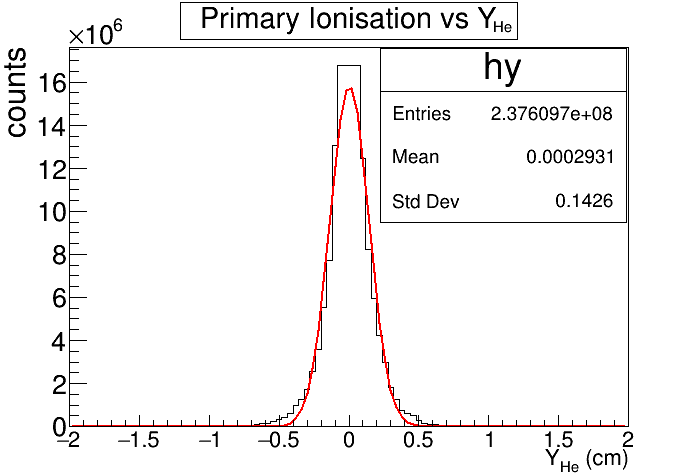}
         \caption{}
         \label{fig2:}
     \end{subfigure}
     % \hspace{10mm}
\begin{subfigure}[b]{0.45\linewidth}
         \centering
         \includegraphics[scale=0.4]{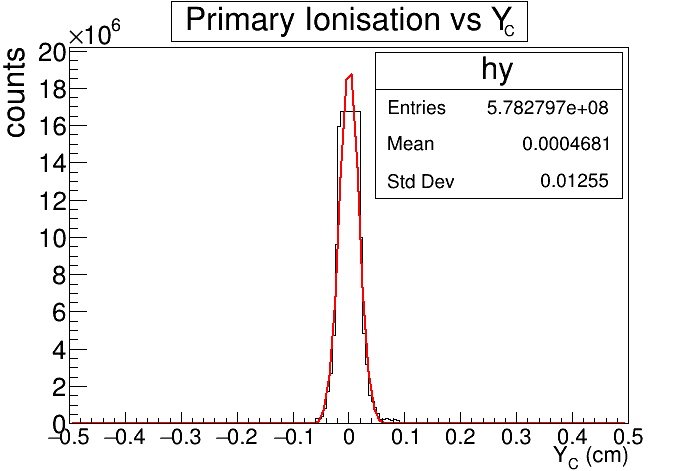}
         \caption{}
         \label{fig2:}
     \end{subfigure}
     \hfill
     % \hspace{20mm}
     \begin{subfigure}[b]{0.45\linewidth}
         \centering
         \includegraphics[scale=0.4]{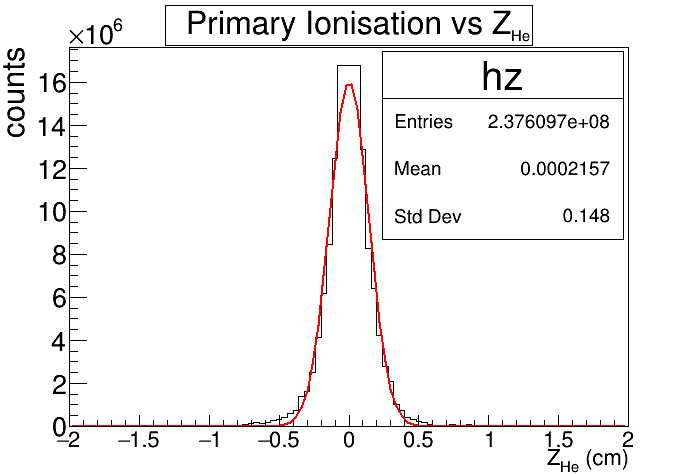}
         \caption{}
         \label{fig3:}
     \end{subfigure}
     % \hspace{30mm}
     \begin{subfigure}[b]{0.45\linewidth}
         \centering
         \includegraphics[scale=0.4]{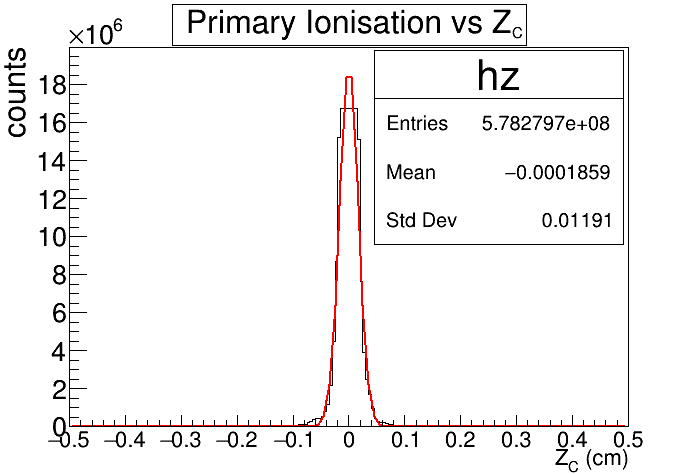}
         \caption{}
         \label{fig3:}
     \end{subfigure}
        \caption{Primary ionization distributions along respective $X$, $Y$ and $Z$-directions, caused by the scattered  $^4He$ of energy 7.6 MeV (a), (c), (e), and $^{12}C$ of energy 17.4 MeV (b), (d), (f) in the drift volume}
        \label{alpha-C}
\end{figure}

The primary ionization distributions caused by the scattered products $^4He$ and $^{12}C$ of energies $7.6~MeV$ and $17.4~MeV$ respectively have been determined by GEANT4 simulation and shown in figure \ref{alpha-C}. The $X$-distributions, where the Bragg peaks are present along the direction of scattered particle $X_{He}$ and $X_C$, have been fitted with the Crystal Ball function, as expressed by equations (\ref{cbf}-\ref{cbf_const}).
\begin{equation}
f(x;\alpha,n,\bar{x},\sigma)=\begin{cases}
exp(\frac{-(x-\bar{x})^2}{2\sigma^2}),& \text{for $\frac{-(x-\bar{x})^2}{\sigma}>-\alpha$} \\
A\times(B-exp(\frac{-(x-\bar{x})^2}{\sigma})),   & \text{for $\frac{-(x-\bar{x})^2}{\sigma} \le -\alpha$}
\end{cases} 
\label{cbf}
\end{equation}
where
\begin{equation}
    A=(\frac{n}{|\alpha|})^n\times exp(-\frac{|\alpha|^2}{2}), 
    B=\frac{n}{|\alpha|} -|\alpha|
    \label{cbf_const}
\end{equation}
The $Y$ (along $Y_{He}$ and $Y_C$) and $Z$ (along $Z_{He}$ and $Z_C$) distributions for both the products have been fitted with the Gaussian function, as has been done previously for the projectile. The fitted parameter for the scattered products have been utilised to form the respective seed clusters that have been subject to propagation through the drift volume. The number of primaries produced by each scattered products, $^4He$ of energy $7.6~MeV$ and $^{12}C$ of energy $17.4~MeV$, are $ 2.37\times10^5$ and $5.43\times10^5$, respectively. 

The table \ref{X_fit} has presented the relevant fitted Crystal Ball function parameters, obtained from equation \ref{cbf}, to describe the ionization along the respective $X$-directions caused by the projectile and scattered particles while those for their $Y$ and $Z$-distributions, fitted with the Gaussian function, have been tabulated in table \ref{Y_Z_fit}. 

\begin{table}[h!]
\caption{Respective $X$-distribution fitted parameters of the projectile and scattered products}
\centering
\begin{tabular}{|c|c|c|c|c|}
\hline
& \multicolumn{4}{c|}{$X$-distribution}\\
\hline
Projectile $^4He$&\multicolumn{4}{c|}{Uniform distribution from $-5$ to $5~cm$} \\
    \hline
     \multirow{2}{*}{Scattered $^4He$}&$\bar{x}$&$\sigma$&$\alpha$&n\\
         \cline{2-5}
& 6.34&0.0827 &0.22 & 0.29 \\
    \hline
Scattered $^{12}C$& 1.808&0.0015 &0.0785 & 0.2938 \\
   \hline
    % etc. ...
    \end{tabular}   
        \label{X_fit}
\end{table}
\begin{table}[h!]
            \caption{Respective $Y$ and $Z$-distribution fitted parameters of the projectile and scattered products}
        \centering
    \begin{tabular}{|c|c|c|c|c|}
    \hline
      & \multicolumn{2}{c|}{$Y$-distribution} & \multicolumn{2}{c|}{$Z$-distribution}  \\
    \hline
     \multirow{2}{*}{Projectile $^4He$}  &$\bar{x}$&$\sigma$ &$\bar{x}$&$\sigma$  \\
    \cline{2-5}
     & 0 & 0.4&0.0&0.4\\
    \hline
Scattered $^4He$&0 & 0.15&0.01&0.128\\
\hline
Scattered $^{12}C$&0 & 0.015&0.0&0.011\\
\hline
    % etc. ...
    \end{tabular}   
        \label{Y_Z_fit}
\end{table}

\subsection{Electric Field}
The drift volume has been segmented appropriately for electric field computation. The maximum and minimum sizes of mesh element used are 0.2 and $0.002~cm$, calibrated for general physics application in COMSOL. The mesh size has been chosen for a reasonable computation time on a workstation with $1.2~Teraflops$ computing performance.  

\subsection{Transport Properties}
\label{TranPro}
\begin{figure}[h!]
     \centering
     \begin{subfigure}[b]{0.45\linewidth}
         \centering
         \includegraphics[width=\textwidth]{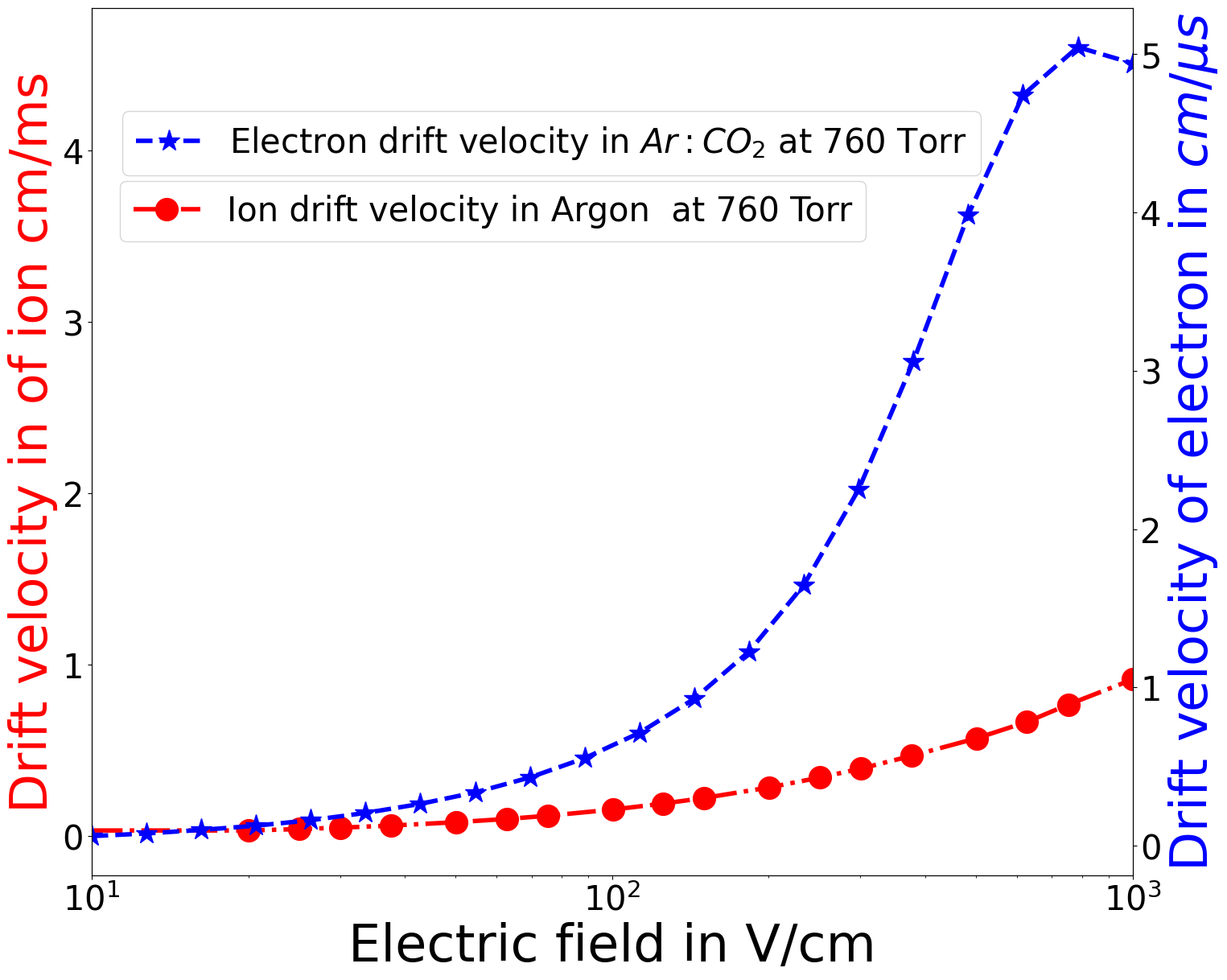}
         \caption{}
         \label{fig:Ardrift}
     \end{subfigure}
     % \hspace{40 mm}
     \begin{subfigure}[b]{0.45\linewidth}
         \centering
         \includegraphics[width=\textwidth]{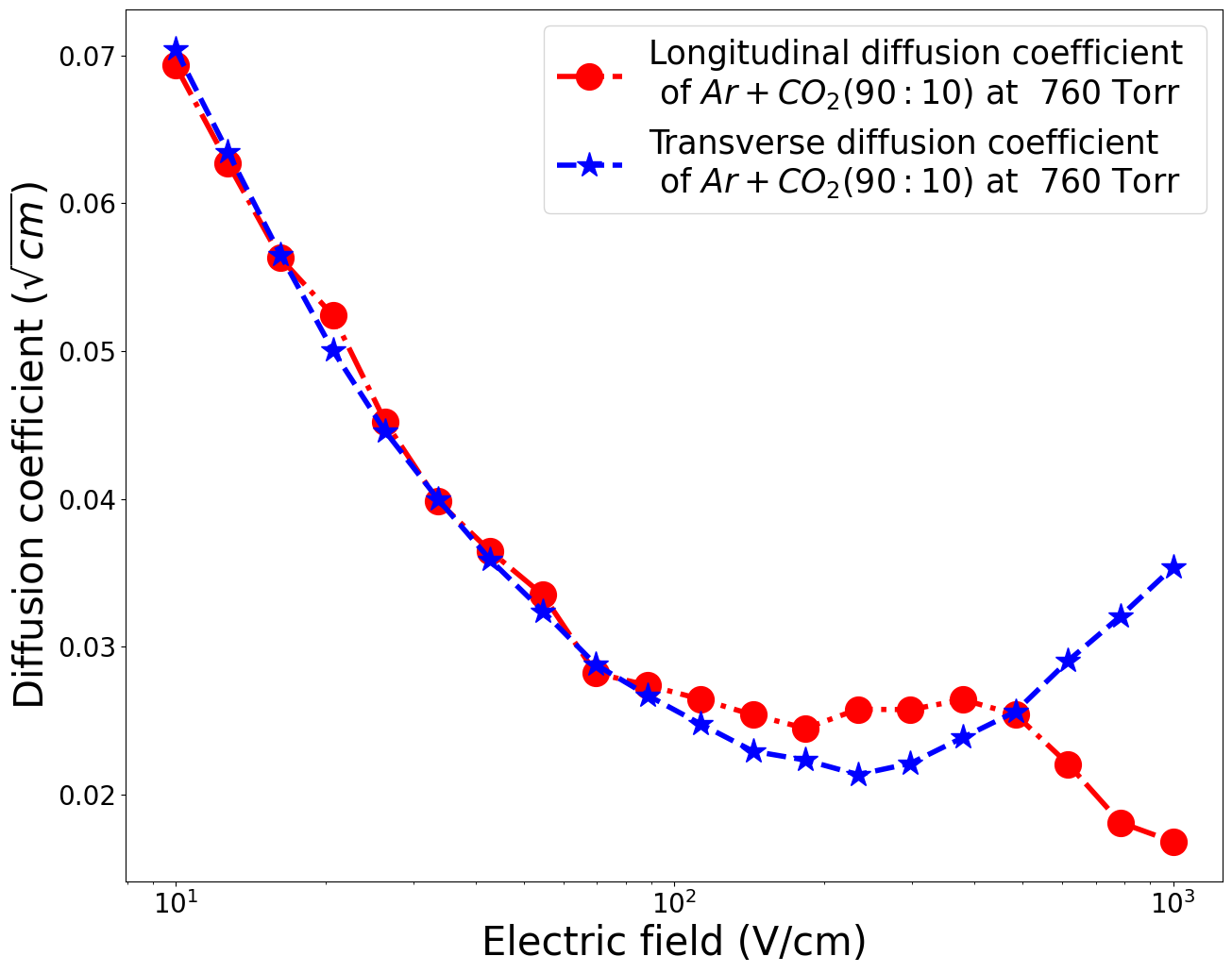}
         \caption{}
         \label{fig:Ardff}
     \end{subfigure}
   \caption{(a) Drift velocity of electrons (from MAGBOLTZ) and ions following \cite{Bas2000}, and (b) diffusion coefficients of electrons in $Ar + CO_2$ (90:10) gas mixture at $760~Torr$ (from MAGBOLTZ)}
        \label{fig:Ar}
\end{figure}

The transport properties, namely the drift velocity and the diffusion coefficients of the electrons in the concerned gas mixture $Ar + CO_2$ (90:10) at $760~Torr$ have been calculated using the MAGBOLTZ. The drift velocity as a function of the electric field has been plotted in figure \ref{fig:Ardrift} whereas that of the Argon ion, as follows from \cite{Bas2000}, has also been depicted. The longitudinal and transverse diffusion coefficients for electrons have been shown in figure \ref{fig:Ardff}.

\section{Beam Induced Space Charge Effect}
\label{BeaSpa}
\begin{figure}[h!]
     \centering
     \begin{subfigure}[b]{0.45\linewidth}
         \centering
         \includegraphics[scale=0.75]{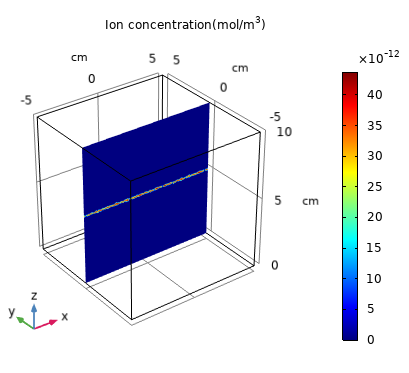}
         \caption{}
         \label{ion_100}
     \end{subfigure}
     \hfill
     \begin{subfigure}[b]{0.45\linewidth}
         \centering
         \includegraphics[scale=0.75]{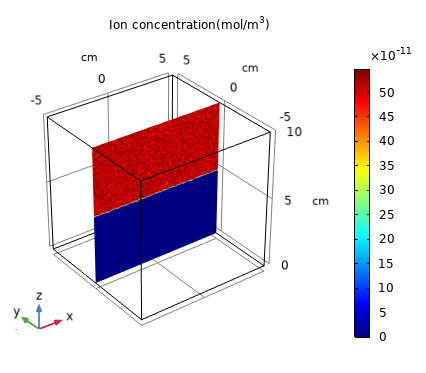}
         \caption{}
         \label{ion_62660}
     \end{subfigure}
        \caption{Ion concentration in X-Z plane at Y = 0 after (a) 10, and (b) 62660 beam events}
        \label{ion}
\end{figure}

The effect of the beam induced space charge on the electric field has been calculated by taking the space charge concentration produced by the $^4He$ projectile beam of energy $25~MeV$ and current $2.3~pA$ into account. It produces a particle rate $0.7186\times10^7$ \textit{pps}. For the sake of simplicity, it has been assumed here that all the beam particles have passed through the device without any interaction.
Owing to the small mobility of the ions and their continuous production by the beam, the space charge concentration on the upper half of the device spreads gradually until it reaches a steady state when the first batch of ions arrives at the cathode plane. In order to obtain a temporal evolution of the space charge concentration, the drift velocity of the Argon ion at $760~Torr$ has been used following the figure \ref{fig:Ardrift}.
After the first beam particle passes through the active volume of the device, the number of ions produced is $1.25\times10^5$, as mentioned in the section \ref{PriIon}. The distance traveled by these ions before the second beam particle comes in is $\Delta t\times v^{ion}_{d} = 798~nm$, where $\Delta t = 140~ns$ represents the beam frequency and $v^{ion}_d = 570~cm/s$ is the drift velocity of ions as per \cite{Bas2000}. Therefore, it can be shown that after 62660 beam particles, the first batch of ions reaches the cathode. A steady state is then achieved as an ion wall of $5~cm$ height (= $62660\times798 nm$) spreading over the range from $ Z =5~cm$ to $ Z =10~cm$ is produced. The ion concentration after the passage of 10 and 62660 beam particles have been shown in figure \ref{ion}. The change in space charge density has been found to vary from $10\times1.25\times10^{5}$ to $62660\times1.25\times10^{5}$ with an increment in the ion wall from $1~mm$ to $5~cm$ in the $XZ$-plane. Here, after 10 events, the height of the ion wall has been considered as $1~mm$ (radius of beam spot) due to the fluctuation of the beam track. 

\begin{figure}[h!]
     \centering
         \centering
         \includegraphics[scale=0.2]{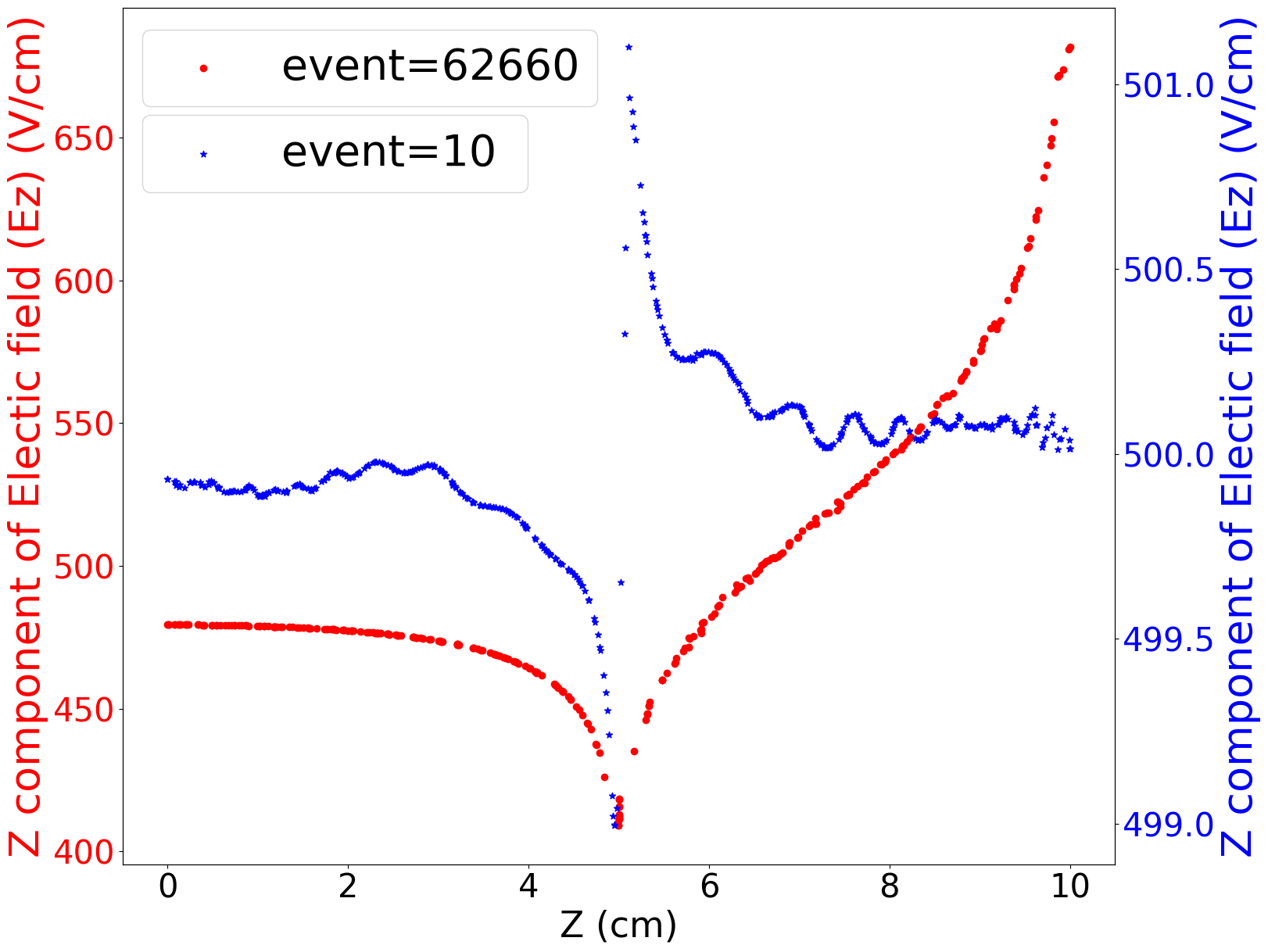}
        \caption{$Z$-component of the electric field along the central axis of the SAT-TPC}
        \label{axis_field}
\end{figure}

The electric field along the central $Z$-axis of the SAT-TPC after passage of 10 and 62660 beam particles has been shown in figure \ref{axis_field}. A drop in the axial field value from $500~V/cm$ to $480~V/cm$ has been observed in the lower half of the drift volume due to the beam induced space charge which should influence the tracking performance of the device for the scattered products. On the other hand, an increasing trend has been found in the upper half.  

\begin{figure}[h!]
 \centering
     \includegraphics[scale=0.6]{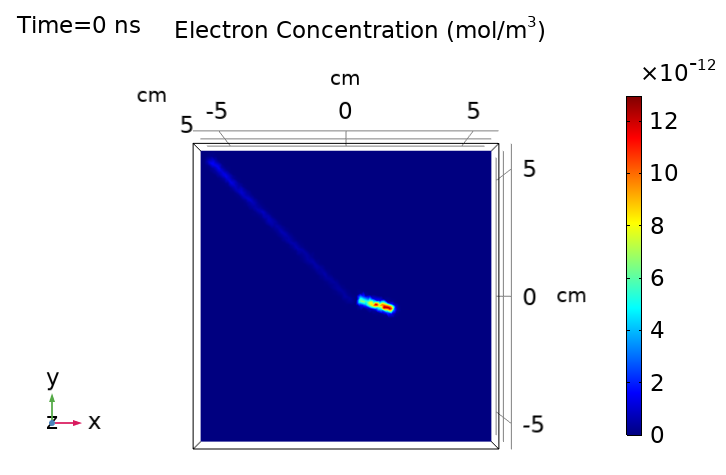}
 \caption{Seed clusters of the scattered products $^4He$ and $^{12}C$ with energy $7.6$, $17.4~MeV$, and angle $135^\circ$, $15.68^\circ$ respectively}
     \label{tracks}
   \end{figure}

\begin{figure}[h!]
     \centering
     \begin{subfigure}[b]{0.95\linewidth}
         \centering
         \includegraphics[scale=0.4]{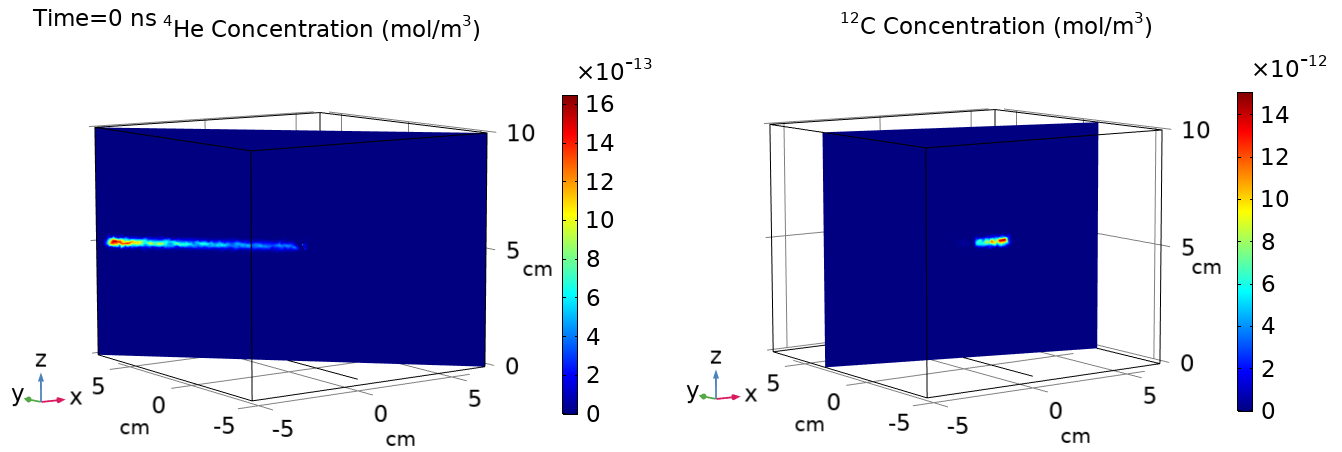}
         \caption{}
         \label{fig1:}
     \end{subfigure}
     \hfill
     \begin{subfigure}[b]{0.95\linewidth}
         \centering
         \includegraphics[scale=0.4]{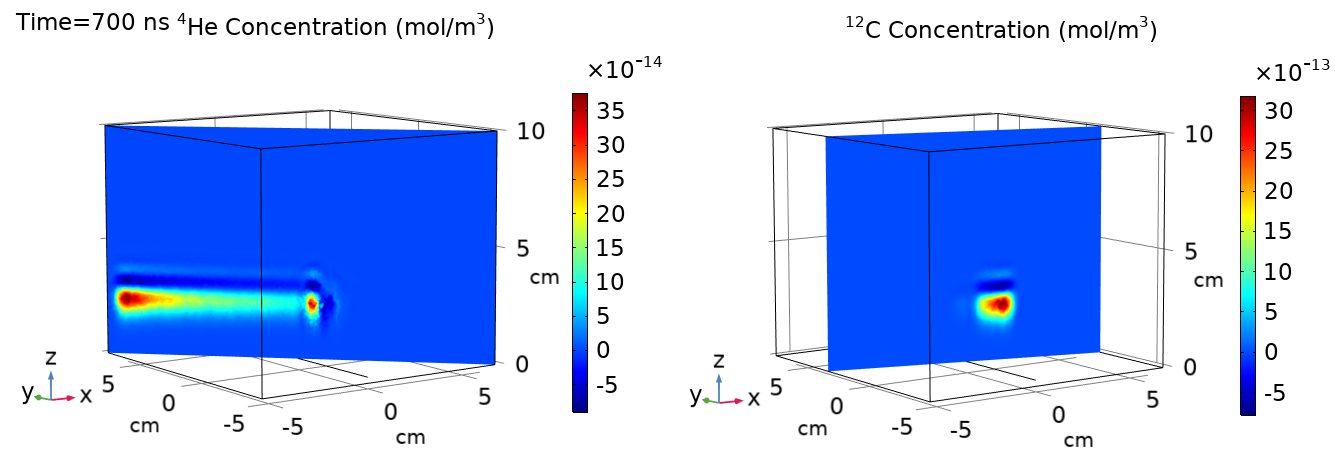}
         \caption{}
         \label{fig2:}
     \end{subfigure}
     \hfill
     \begin{subfigure}[b]{0.95\linewidth}
         \centering
         \includegraphics[scale=0.4]{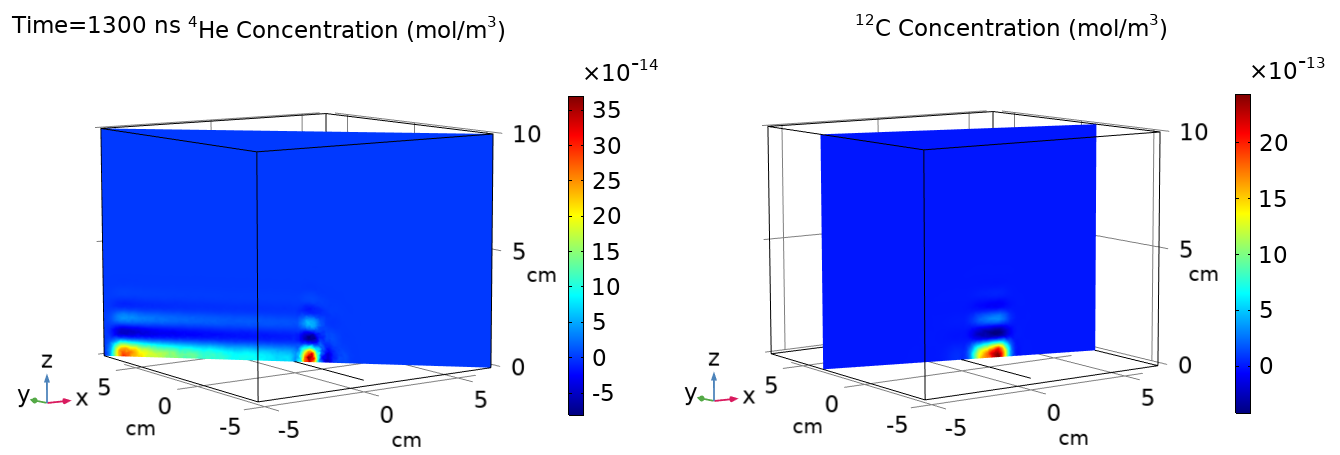}
         \caption{}
         \label{fig3:}
     \end{subfigure}
        \caption{Concentration of primary electrons in scattered particle tracks in the drift volume at (a) $Time = 0~ns$, (b) $Time =700~ns$, and (c) $Time =1300~ns$}
        \label{transport}
\end{figure}

The seed clusters of the scattered particles $^4He$ of energy $7.6~MeV$ and $^{12}C$ of energy $17.4~MeV$ have been constructed by using the parameters extracted from their primary ionization distributions. The clusters have been placed in the $XY$-plane at $Z=5~cm$ and with their specific angle of scattering $135^\circ$ and $15.68^\circ$, as shown in figure \ref{tracks}.
It may be noted that the longer $^4He$ track cluster at ${135}^\circ$ angle has less electron concentration while the shorter $^{12}C$ track at $15.68^\circ$ has more. Subsequently, the primary electrons of the respective clusters have been propagated following the procedure discussed earlier. The electron concentration has been calculated at different instants, $Time  = 0, 700, 1300~ns$ for both the track clusters, taking the electric field updated with the temporal evolution of the space charge into consideration. The results have been illustrated in figure \ref{transport}.

\begin{figure}[h!]
     \centering
     \begin{subfigure}[b]{0.45\linewidth}
         \centering
         \includegraphics[scale=0.18]{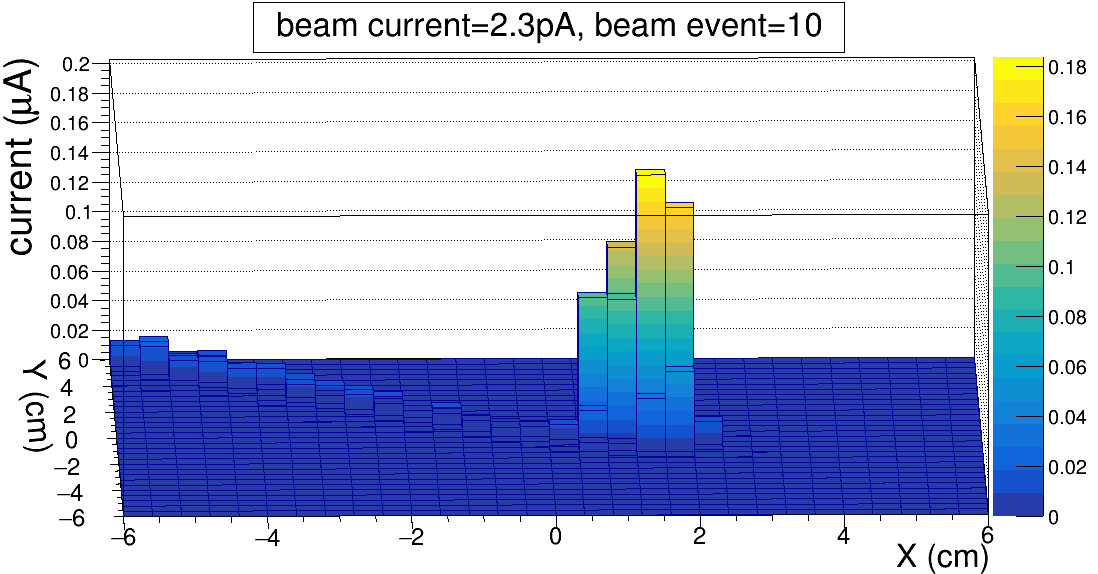}
         \caption{}
         \label{track_100_a}
     \end{subfigure}
     \hfill
     \begin{subfigure}[b]{0.45\linewidth}
         \centering
         \includegraphics[scale=0.18]{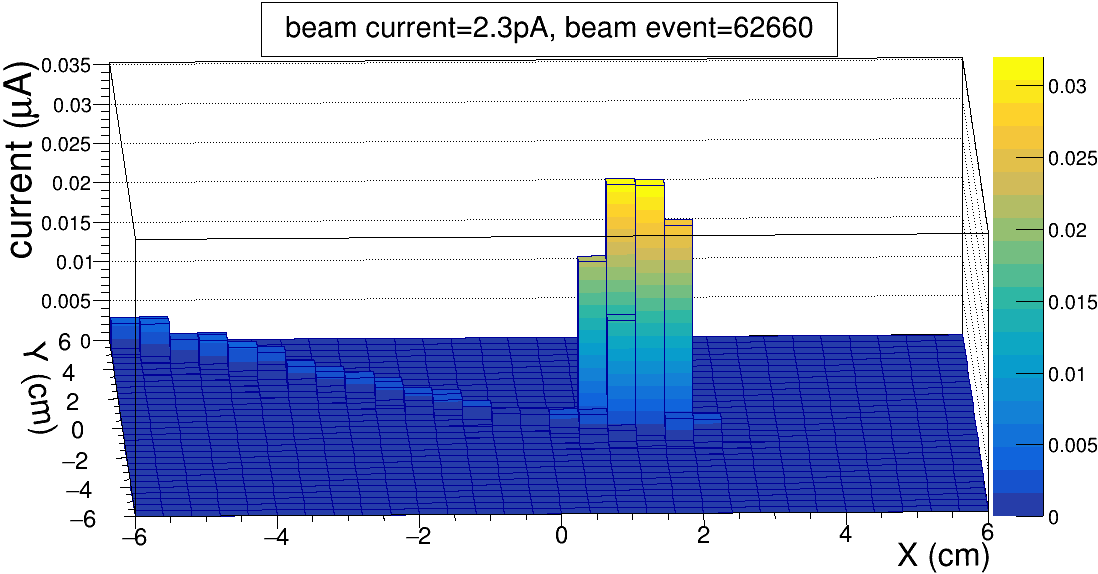}
         \caption{}
         \label{track_100}
     \end{subfigure}
        \caption{Current, generated by scattered particle tracks at anode plane with segmentation $4~mm\times4~mm$ after (a) 10, and (b) 62660 beam events}
        \label{lego_track}
\end{figure}

\begin{figure}[h!]
    \centering
     \begin{subfigure}[b]{0.45\linewidth}
       \centering
       \includegraphics[scale=0.4]{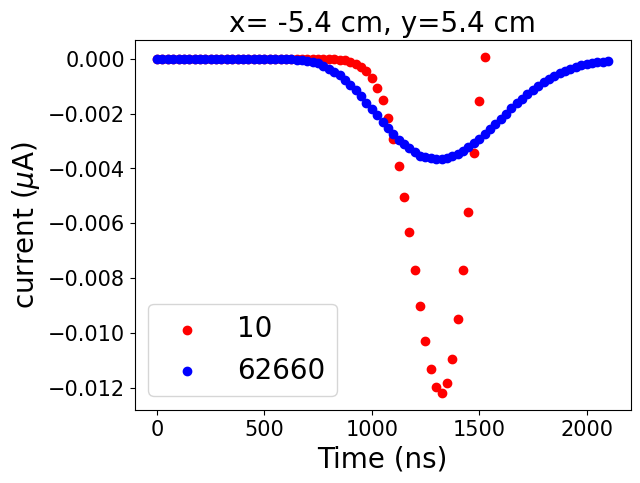}
       \caption{}
       \label{4He}
     \end{subfigure}
     % \hspace{20mm}
     \begin{subfigure}[b]{0.45\linewidth}
        \centering
       \includegraphics[scale=0.4]{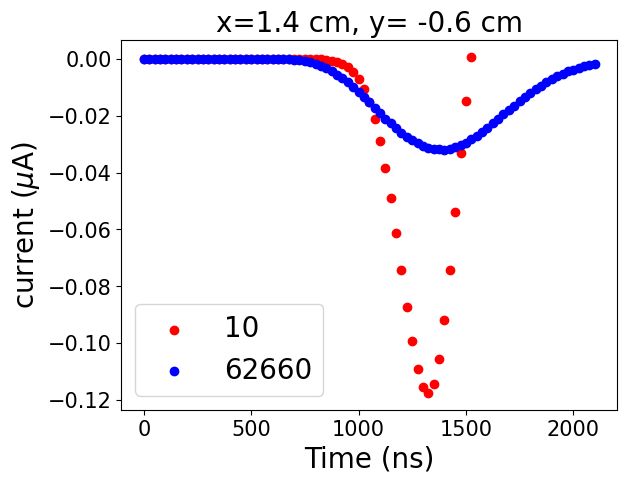}
       \caption{}
       \label{12C}
     \end{subfigure}
    \caption{Current on anode readout pads, located at (a) $ X = -5.4~cm$, $Y = 5.4~cm$ on $^4He$ track, and (b) $X = 1.4~cm$, $Y = -0.6~cm$ on $^{12}C$ track for beam event numbers 10 and 62660}
    \label{anode_current}
\end{figure}

Following their propagation, the current signals produced by the scattered track clusters of $^4He$ and $^{12}C$ on the anode plane, segmented in $4~mm\times4~mm$ readout pads, have been computed. The 3D histogram of the same has been plotted in figure \ref{lego_track} for the space charge produced by beam events 10 and 62660.  It may be noted by comparing the current values around the Bragg's peaks of the scattered particles that the space charge produced by the 62660 beam events has caused a reduction in the same.
To make a quantitative estimate, the current values, produced on specific readout pads, lying under the Bragg peak of the tracks of two scattered products, have been plotted in figure \ref{anode_current} for two different numbers of beam events. It has shown that the current on the given readout pad has reduced by $25-30\%$ approximately due to the presence of beam induced space charge.   

\section{Beam Current Optimization}
\begin{figure}[h!]
     \centering
          \begin{subfigure}[b]{0.45\linewidth}
         \centering
         \includegraphics[scale=0.18]{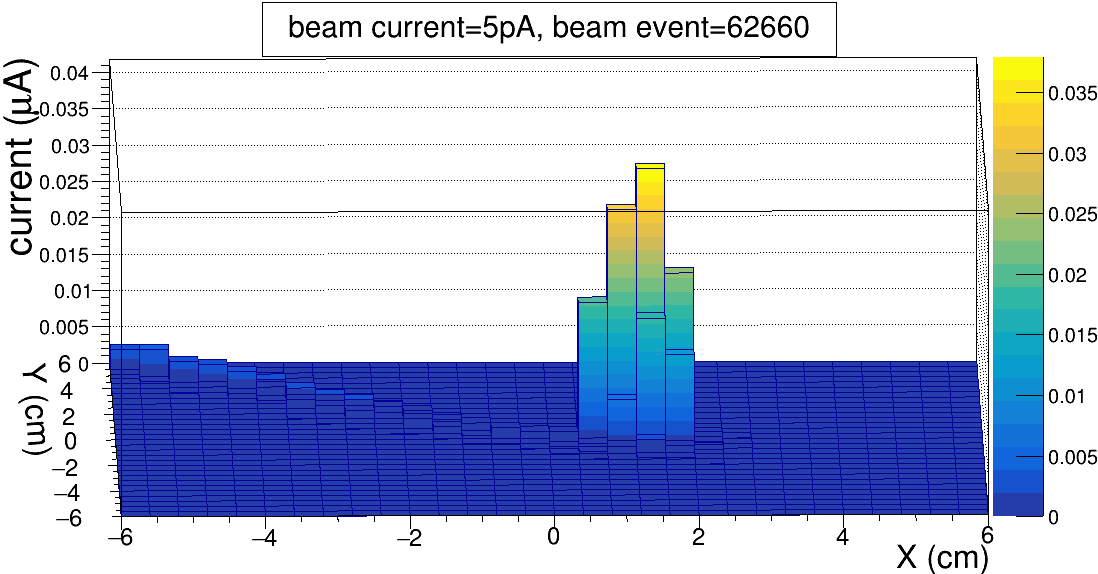}
         \caption{}
         \label{fig1:}
     \end{subfigure}
     % \hspace{6 mm}
     \hfill
     \begin{subfigure}[b]{0.45\linewidth}
         \centering
         \includegraphics[scale=0.18]{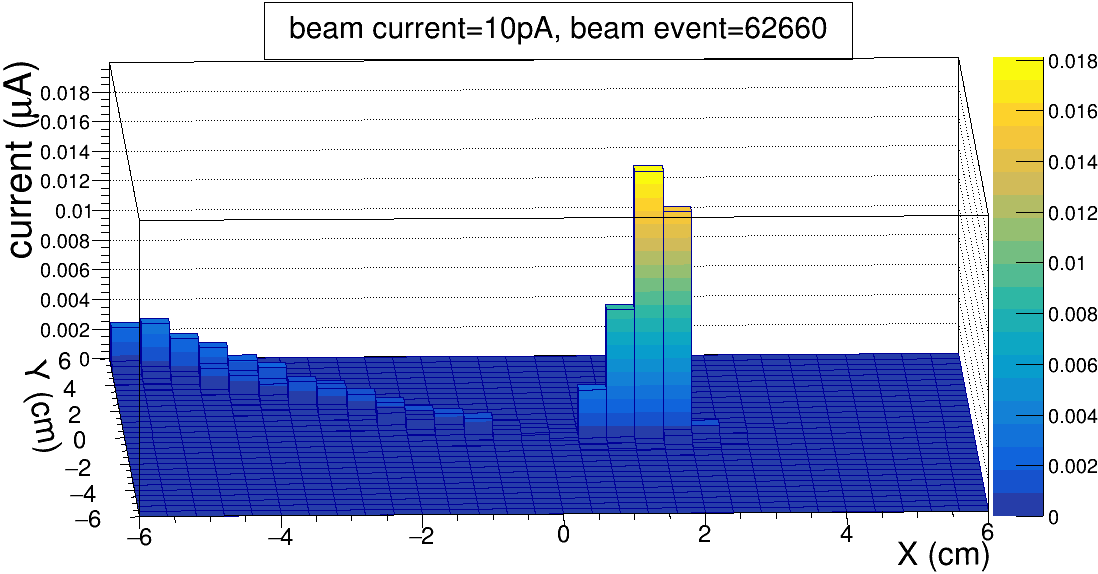}
         \caption{}
         \label{fig1:}
     \end{subfigure}
     % \hspace{1mm}
          \hfill
     \begin{subfigure}[b]{0.45\linewidth}
         \centering
         \includegraphics[scale=0.18]{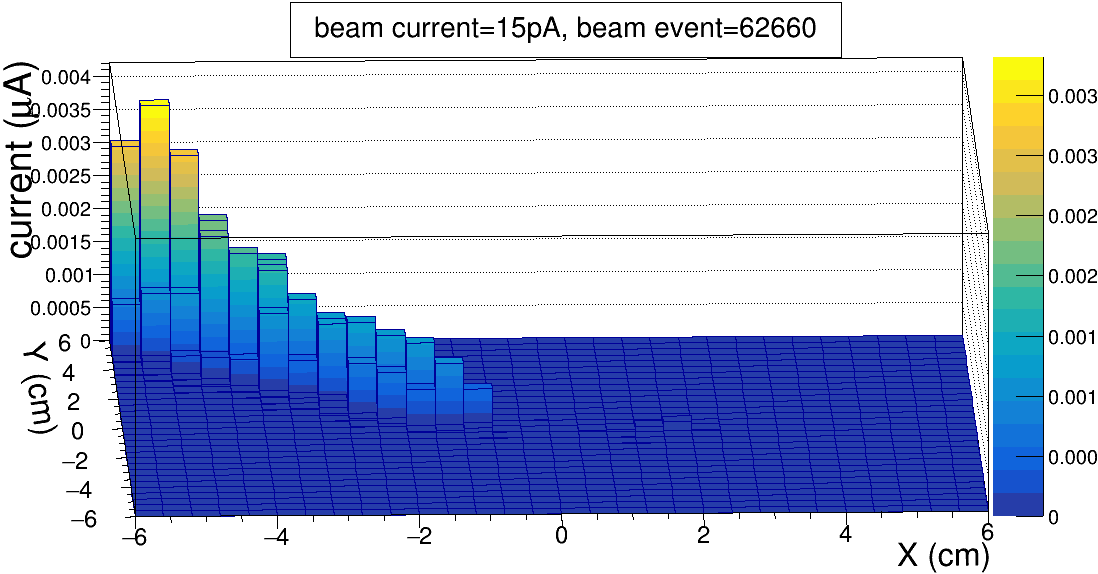}
         \caption{}
         \label{fig2:}
     \end{subfigure}
%     \hspace{12 mm}
 %    \begin{subfigure}[b]{0.2\textwidth}
 %        \centering
  %       \includegraphics[scale=0.12]{TPC_anode_signal_62660_lego_20pA.png}
 %        \caption{}
 %        \label{fig3:}
 %    \end{subfigure}
        \caption{Anode current produced by scattered particle tracks after 62660 beam events for beam currents (a) $5~pA$, (b) $10~pA$, and (c) $15~pA$}
        \label{sc_tracks}
\end{figure}

A numerical investigation has been carried out to study the effect of variation in the beam current from $2.3~pA$ to $15~pA$ and the resultant space charge effect. In table \ref{ion_electron}, the corresponding steady state space charge concentrations have been tabulated for each beam current produced after 62660 beam events.
\begin{table}[h!]
\caption{Steady state space charge concentration at each beam current}
\centering
\begin{tabular}{|m{2cm}|m{2cm}|m{3cm}|}
\hline
  Current (\textit{pA}) & \textit{pps}  & Ion concentration  \\ \hline
2.3 & 0.718$\times10^7$ & 7.56$\times10^9$  \\ \hline
5  & 1.563$\times10^7$  & 1.655$\times10^{10}$  \\ \hline
10 & 3.126$\times10^7$ & 3.446$\times10^{10}$ \\ \hline
15 & 4.889$\times10^7$ & 4.957$\times10^{10}$ \\ \hline
%20 & 6.25$\times10^7$  & 6.892$\times10^{10}$ \\ \hline
\end{tabular}
\label{ion_electron}
\end{table}
In figure \ref{sc_tracks}, the current produced by the scattered particle tracks after 62660 beam events for each beam current has been plotted. It can be observed from the 3D plots that the space charge effect has become substantial beyond $10~pA$ so that the scattered $^{12}C$ track could not be recorded.

Based on the simulated results along with the basic assumption of no interaction of the beam particles with the medium till the steady state has been reached, it can be stated  that the optimum range of the beam current should be less than $10~pA$. Beyond this range, there will be significant track distortion due to the space charge effect which may lead to difficulty in track recording.  From the Monte Carlo simulation, it has been found that 2.1\% events can undergo scattering for the given case. This will increase the optimum limit of the beam current, but not significantly. 

\section{ Anode Segmentation Optimisation}
\begin{figure}[htb]
     \centering
     \begin{subfigure}[b]{0.45\linewidth}
         \centering
         \includegraphics[width=\textwidth]{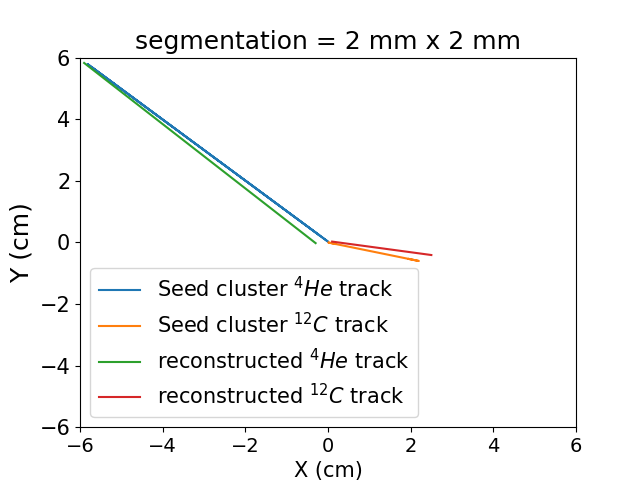}
         \caption{}
         \label{fig0:}
     \end{subfigure}
     % \hfill
     \begin{subfigure}[b]{0.45\linewidth}
         \centering
         \includegraphics[width=\textwidth]{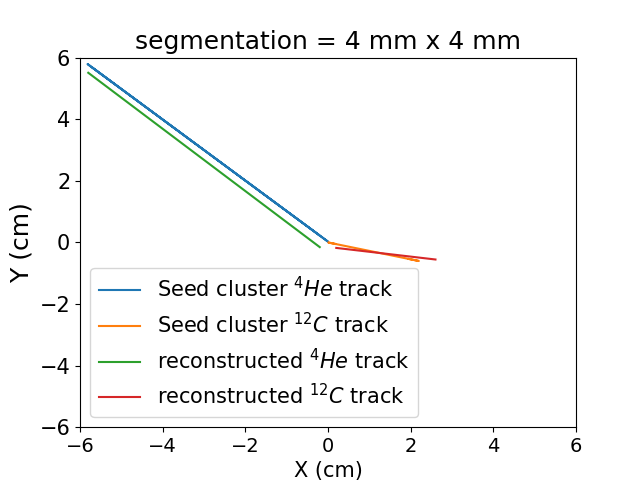}
         \caption{}
         \label{fig1:}
     \end{subfigure}
      % \hspace{10 mm}
     \begin{subfigure}[b]{0.45\linewidth}
         \centering
         \includegraphics[width=\textwidth]{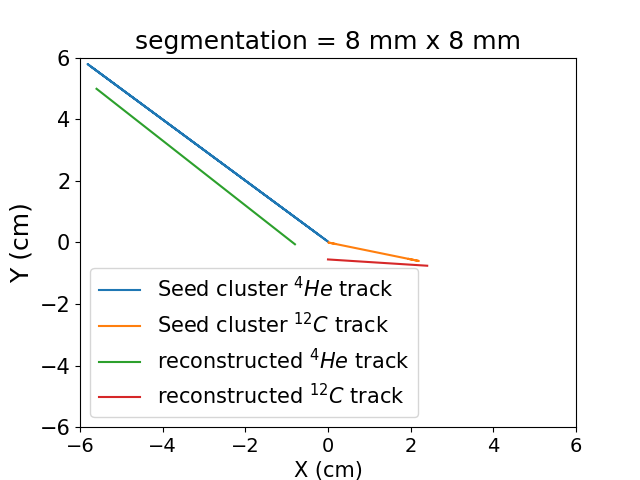}
         \caption{}
         \label{fig2:}
     \end{subfigure}
        \caption{Scattered particle tracks (seed cluster and reconstructed) after 62660 beam events with beam current $5~pA$, and readout segmentation (a) $2~mm\times2~mm$, (b) $4~mm\times4~mm$, and (c) $8~mm\times8~mm$}
        \label{seg}
\end{figure}

%The effect of anode segmentation on the angular resolution of the reconstructed tracks of scattered particles has been investigated as well. 
The current signal, generated by the scattered particle tracks on the anode plane has been studied with different segmentation schemes of the anode plane. It has been made from $2~mm\times2~mm$, $4~mm\times4~mm$, and $8~mm\times8~mm$. The simulation has been carried out with beam current $5~pA$ for 62660 beam events when the steady state is achieved. The tracks reconstructed from the current signals, of the two scattered particles, have been compared to the original track of their seed clusters. 
The projected tracks has been reconstructed using linear regression with SciPy library \cite{scipy2001} of the readout positions produced by the center of gravity method. In the calculation, amplitude of the current has been used as a weight factor. 
The results for the three cases of the segmentation schemes, $2~mm\times2~mm$, $4~mm\times4~mm$, and $8~mm\times8~mm$, have been illustrated in figure \ref{seg}. 

The angle, obtained from the reconstructed tracks of the scattered particles, $^4He$ and $^{12}C$, for different segmentation schemes, have been tabulated in table \ref{table_angles} to compare with the actual angles, mentioned in table \ref{tab:my_label}. 
\begin{table}[h!]
\caption{Effect of anode segmentation on reconstructed angles}
\centering
\begin{tabular}{|m{3cm}|m{2cm}|m{2cm}|m{2cm}|m{2
cm}|}
\hline
& {$2~mm\times2~mm$}&{$4~mm\times4~mm$}&{$8~mm\times8~mm$}& seed cluster\\
\hline
% \cline{2-6}
{$^4He$ track angle}&$133.75^\circ\pm0.36$ & $134.13^\circ\pm1.54$&$133.56^\circ\pm1.41$&$135.33^\circ\pm0.3$ \\
\hline
{$^{12}C$ track angle}&$10.3^\circ\pm1.34$&$6.56^\circ\pm0.98$&$4.85^\circ\pm1$&$15.27^\circ\pm1.81 $\\
\hline
\end{tabular}
 \label{table_angles}
\end{table}
It can be observed from the tabulated data that the effect of segmentation of anode  is much more pronounced in case of $^{12}C$ track than the $^4He$ one, due to its short length. It is obvious from the plots and the angle data that the segmentation $2~mm\times2~mm$ provides better resolution with respect to other choices. However, in order to compromise between the quality of data and cost effective fabrication of the device, the segmentation scheme of $4~mm\times4~mm$ may be opted for which the 3D plots of current signals of the scattered tracks have been shown in the previous figure \ref{lego_track}.

\section{Summary and Conclusion}
\label{SumCon}
The present hydrodynamic model, which is computationally less expensive in comparison to particle model, has been found to perform satisfactorily in emulating the device dynamics of an AT-TPC. It has produced nice agreement when compared to the particle model in explaining the experimental observations of scattered particle track distortion due to beam induced space charge in a low energy nuclear physics experiment. The hydrodynamic model has been utilized for optimization of operational limit of beam current and anode segmentation of the proposed SAT-TPC prototype to study non-relativistic elastic scattering of $^4He$ of energy $25~MeV$ with active gaseous target $^{12}C$. On the basis of the space charge effect produced by the projectile beam along with an assumption of no interaction of beam particles till the steady state of ion concentration is reached, it is suggested that the beam current should be restricted to below $10~pA$, where the signal amplitude for the scattered particle tracks has been found to reduce by $25-30\%$. The numerical investigation on the anode segmentation to optimize its dimension has led to a choice of $4~mm\times4~mm$ for satisfactory tracking performance along with cost effective production of the SAT-TPC prototype. 
The effect of several other physical processes, like the ion backflow from the multiplication stage and the space charge produced by the scattered particles, are planned to be explored in future studies. The scope of possible interaction of the projectile beam with other nuclei present in the active gas mixture is another important aspect to be considered to envisage the practical scenario.  

\section*{Acknowledgements}
The authors would like to acknowledge the support provided their respective institutions. They are thankful to Dr. Tilak Ghosh of Variable Energy Cyclotron Centre, Kolkata, India, for his valuable advises.

% \bibliographystyle{elsarticle-harv}

% \bibliography{Bib.v5}

\begin{thebibliography}{99}

\bibitem{Nyg1974}
D. R. Nygren,
\emph{Proposal to Investigate the feasibility of a Novel Concept in Particle Detection},
\href{https://inspirehep.net/literature/1365360}{\emph{LBL internal report} {} (1974), 22-70.}

\bibitem{Mar1978}
Jay N. Marx and David R. Nygren,
\emph{The Time Projection Chamber},
\href{https://pubs.aip.org/physicstoday/article-abstract/31/10/46/431734/The-Time-Projection-ChamberBy-combining-the?redirectedFrom=fulltext}{\emph{Physics Today},{\bf vol 861}, (1978) 46-53.}

\bibitem{Att2009}
D. Attie,
\emph{TPC review},
 \href{https://www.sciencedirect.com/science/article/pii/S0168900208011996}{\emph{Nuclear Inst. and Methods in Physics Research A},{\bf vol 598},(2009) 89-93.}

\bibitem{Hil2010}
H. J. Hilke,
\emph{Time projection chambers}, \href{https://iopscience.iop.org/article/10.1088/0034-4885/73/11/116201}{
\emph{Reports on Progress in Phyics},{\bf vol 73},(2010),116201-116237.}

\bibitem{Die2018}
Diego González-Díaz and Francesc Monrabal and Sebastien Murphy,
\emph{Gaseous and dual-phase time projection chambers for imaging rare processes},
\href{https://www.sciencedirect.com/science/article/pii/S0168900217309944?via%3Dihub}{\emph{Nuclear Inst. and Methods in Physics Research A},{\bf vol 868},(2018), 200-255.}

\bibitem{Ayy2018}
Y. Ayyad and D. Bazin and S. Beceiro-Novo and M. Cortesi,
\emph{Physics and technology of time projection chambers as active targets},
\href{https://link.springer.com/article/10.1140/epja/i2018-12557-7}{\emph{The European Physical Journal A},{\bf vol 54},(2018) 181.}

\bibitem{Baz2020}
D. Bazin and T. Ahn and Y. Ayyad and S. Beceiro-Novo and A. O. Machhiavelli,
\emph{Low energy nuclear physics with active target time projection chambers},
\href{https://www.sciencedirect.com/science/article/abs/pii/S0146641020300375}{\emph{Progress in Particle and Nuclear Physics},{\bf vol 114},(2020),103-790.}

\bibitem{Ran2019}
Jaspreet S Randhawa and M. Cortesi and Y. Ayyad and W. Mittig
\emph{Beam induced space-charge effects in Time Projection Chambers
in low-energy nuclear physics experiments},
\href{https://www.sciencedirect.com/science/article/pii/S0168900219312641}{\emph{Nuclear Inst. and Methods in Physics Research, A},{\bf vol 948},(2019) 162830.}
\bibitem{deb2017}
Deb Sankar Bhattacharya and Purba Bhattacharya and Prasant Kumar Rout and Supratik Mukhopadhyay and Sudeb Bhattacharya and Nayana Majumdar and Sandip Sarkar and Paul Colas and David Atti’e and Serguei Ganjour and Aparajita Bhattacharya,
\emph{Experimental and numerical simulation of a TPC like set up for the measurement of ion backflow},
\href{https://www.sciencedirect.com/science/article/pii/S0168900217304837}{\emph{Nuclear Inst. and Methods in Physics Research A},{\bf vol 861},(2017) 64-70.}

\bibitem{Purba2017}
Purba Bhattacharya and Bedangadas Mohanty and Supratik Mukhopadhyay and Nayana Majumdar,
 Hugo Natal da Luzc
\emph{3D simulation of electron and ion transmission of GEM-based detectors},
\href{https://www.sciencedirect.com/science/article/pii/S0168900217307076}{\emph{Nuclear Inst. and Methods in Physics Research, A},{\bf vol },(2017) 64-72.}


\bibitem{Dat2021}
Jaydeep Datta and S. Tripathy and N. Majumdar and S. Mukhopadhyay
\emph{Numerical qualification of eco-friendly gas mixtures for avalanche-mode operation of resistive plate chambers in INO-ICAL},
\href{https://iopscience.iop.org/article/10.1088/1748-0221/16/07/P07012}{\emph{Journal of Instrumentation},{\bf vol 17},(2021), 07-012.}

\bibitem{Dat2020}
Jaydeep Datta and S. Tripathy and N. Majumdar and S. Mukhopadhyay,
\emph{Study of Streamer development in Resistive Plate Chamber},
\href{https://iopscience.iop.org/article/10.1088/1748-0221/15/12/C12006}{\emph{Journal of Instrumentation},{\bf vol 16},(2021), P02018.}

\bibitem{Rou2021}
Prasant Kumar Rout and J. Datta and P. Roy and P. Bhattacharya,
\emph{Fast simulation of avalanche and streamer in GEM detector using hydrodynamic approach},
\href{https://iopscience.iop.org/article/10.1088/1748-0221/16/02/P02018}{\emph{Journal of Instrumentation},{\bf vol 16},(2021), P02018.}

\bibitem{Rou2021disprob}
Prasant Kumar Rout and R. Kanishka and J. Datta and P. Roy and P. Bhattacharya and S. Mukhopadhyay and N. Majumdar and S. Sarkar
\emph{Numerical estimation of discharge probability in GEM-based detectors},
\href{https://iopscience.iop.org/article/10.1088/1748-0221/16/09/P09001/pdf}{\emph{Journal of Instrumentation},{\bf vol 16},(2021), P09001}.

\bibitem{Com}
COMSOL Multiphysics
\emph{\href{https://www.comsol.co.in/comsol-multiphysics}{https://www.comsol.co.in/comsol-multiphysics}}
.

\bibitem{Gea2003}
S. Agostinelli and J. Allison and K. Amako and J. Apostolakis
\emph{GEANT4–a simulation toolkit},
\href{https://www.sciencedirect.com/science/article/pii/S0168900203013688?via%3Dihub}{\emph{Nuclear Inst. and Methods in Physics Research, A},{\bf vol 506},(2003) 250-303}.
\bibitem{Mag}
S.F. Biagi
\emph{Monte Carlo simulation of electron drift and diffusion in counting gases under the influence of electric and magnetic fields},
\href{https://www.sciencedirect.com/science/article/pii/S0168900298012339?via%3Dihub}{\emph{Nuclear Inst. and Methods in Physics Research, A},{\bf vol 421},(1999),234-240.}
\bibitem{veenhof1998}
R. Veenhof
\emph{GARFIELD, recent developments},\href{https://www.sciencedirect.com/science/article/pii/S0168900298008511}{\emph{Nuclear Inst. and Methods in Physics Research, A},{\bf vol 419},(1998), 726–730.}

\bibitem{Bas2000}
E. Basurto and J. de Urquijo and I. Alvarez and C. Cisneros,
\emph{Mobility of He+,Ne+,Ar+,N+2,O+2,and CO+2 in their parent gas}
\href{https://journals.aps.org/pre/abstract/10.1103/PhysRevE.61.3053}{,\emph{Physical Review E},{\bf vol 61}(2000),3053.}
\bibitem{scipy2001}
Eric Jones and Travis Oliphant and Pearu Peterson and others
\emph{SciPy: Open source scientific tools for Python},
,(2001), {\href{http://www.scipy.org/}{http://www.scipy.org/}}.

\end{thebibliography}


%% For example:

%% \bibitem[Aladro et al.(2015)]{Aladro15} Aladro, R., Martín, S., Riquelme, D., et al. 2015, \aas, 579, A101

%%\end{thebibliography}

\end{document}